\documentclass{aa}  
\usepackage{graphicx}
\usepackage{txfonts}
\usepackage[]{hyperref}

\usepackage{tablefootnote}
\usepackage{subfigure}
\usepackage[symbol]{footmisc}

\hypersetup{
    colorlinks=true,
    citecolor=blue,
    linkcolor=blue,
    filecolor=magenta,      
    urlcolor=cyan,
}

\usepackage{xcolor}

\usepackage{natbib}
\bibpunct{(}{)}{;}{a}{}{,} 

\usepackage{graphicx}
\usepackage[utf8]{inputenc}

\usepackage[english]{babel}
\usepackage{amsmath}
\usepackage{wasysym}
\usepackage{tabularx}
\usepackage{multirow}
\usepackage{booktabs}
\usepackage{amssymb}
\usepackage{rotating}

\makeatletter
\let\linenumbers\relax
\let\endlinenumbers\relax
\makeatother

\begin{document}

\title{Galaxy cluster virial quantities from extrapolating strong lensing mass profiles}
\author{Enrico Maraboli\inst{1}
  \and Claudio Grillo\inst{1,2}
     \and Pietro Bergamini\inst{1,3}
     \and Carlo Giocoli\inst{3,4}}

\institute{Dipartimento di Fisica, Università degli Studi di Milano, Via Celoria 16, I-20133 Milano, Italy
  \and INAF - IASF Milano, via A. Corti 12, I-20133 Milano, Italy
  \and INAF – OAS, Osservatorio di Astrofisica e Scienza dello Spazio di Bologna, via Gobetti 93/3, I-40129 Bologna, Italy
  \and INFN – Sezione di Bologna, viale Berti Pichat 6/2, I-40127 Bologna, Italy
}

\abstract{We study the radial total mass profiles of nine massive galaxy clusters ($M_\mathrm{200c}>5\times10^{14}$ M$_\odot$) in the redshift range $0.2 < z < 0.9$. These clusters were observed as part of the CLASH, HFF, BUFFALO, and CLASH-VLT programs, that provided high-quality photometric and spectroscopic data. Additional high-resolution spectroscopic data were obtained with the integral-field spectrograph MUSE at the VLT. Our research takes advantage of strong lensing analyses that rely on deep panchromatic and spectroscopic measurements. From these data, we measure the projected total mass profiles of each galaxy cluster in our sample. 
We fit these mass profiles with simple, one-component, spherically symmetric mass models including the Navarro-Frenk-White (NFW), non-singular isothermal sphere, dual pseudo isothermal ellipsoidal, beta model, and Hernquist profiles. We perform a Bayesian analysis to sample the posterior probability distributions of the free parameters of the models. We find that the NFW, Hernquist, and beta models are the most suitable profiles to fit the measured projected cluster total mass profiles. Moreover, we test the robustness of our results by changing the region in which we perform the fits: we slightly modify the center of the projected mass profiles and the radial range of the considered region. We employ the results obtained with the Hernquist profile to compare our total mass estimates ($M_\mathrm{H}^\mathrm{tot} = M_\mathrm{H} (r\rightarrow + \infty)$), with the $M_\mathrm{200c}$ values from weak lensing studies. Through this analysis, we find scaling relations between $M_\mathrm{H}^\mathrm{tot}$ and $M_\mathrm{200c}$ and the value of the scale radius, $r_\mathrm{S}$, and $R_\mathrm{200c}$. Interestingly, we also find that the $M_\mathrm{200c}$ values, obtained by extrapolating the fitted total mass profiles, are very close to the weak lensing results. This feature can be exploited in future studies on clusters and cosmology, as it provides an easy way to infer galaxy cluster virial masses.}

\maketitle

\section{Introduction} \label{sec.intro}

Galaxy clusters play a crucial role both in astrophysics and cosmology, making them among the most interesting physics laboratories to study the evolution and the current state of the Universe. For this reason, they were the targets of many recent observational campaigns, carried on by facilities like the Hubble space telescope (HST), the Spitzer space telescope (SST), the Röntgensatellit (ROSAT), and the James Webb Space Telescope (JWST) (e.g., \citealt{Postman_2012_CLASH}; \citealt{Mann_2012_MACS}; \citealt{Lotz_2017_HFF}; \citealt{Bezanson_2024_UNCOVER}; \citealt{Treu_2022_GLASSJWST}). In addition to these space-based observations, ground-based complementary follow-up programs have been set to complete the photometric and spectroscopic datasets (see e.g. \citealt{Rosati_2014_CLASH}). The general purpose of this intense data collection is to provide the most accurate description of galaxy clusters, by creating a detailed multi-wavelength view on such structures, and on a wide range of scales.
One of the most interesting features of galaxy clusters is their mass composition, which includes not only the stellar component ($\leq 5\%$ of total mass), but also dark matter (DM, 80\%-85\% of total mass) and hot intracluster gas (a.k.a. intracluster medium, ICM, 10\%-15\% of total mass). Due to their dynamical equilibrium status, the analysis of the mass components of the galaxy clusters makes them act as very sensitive and effective cosmological probes \citep{Kravtsov_2012_clusterformation}. By studying the galaxy cluster matter composition and the distribution of the different mass components, we can provide a fair test to the currently accepted cosmological model, through testing the dark matter accretion theories or the hierarchical paradigm growth of large-scale structures, and constrain the normalized value of the baryonic density of the Universe, $\Omega_\mathrm{B}$ \citep{Bahcall_2014_DMtests}. Moreover, the cosmological simulations parameters can be tuned better if based on accurate studies on mass composition, to provide further checks and insights into the current cosmological theories \citep{Huss_1999_cosmogonies}. 

This work aims to study the total mass distribution of galaxy clusters starting from the results obtained by strong lensing (SL) that exploit the new results which come from the aforementioned campaigns. In this paper we analyze nine massive (total mass $M_{200\mathrm{c}}>5\times10^{14} M_\odot$) galaxy clusters in the redshift range $0.2 < z < 0.9$, by fitting the projected radial mass profiles of the total mass with a set of single-component, spherical models. We base our work on robust mass measurements obtained from the most recent SL studies (\citealt{Caminha_2016_AS1063}; \citealt{Caminha_2017_M0416}; \citealt{Caminha_2017_M1206}; \citealt{Bonamigo_2017_M0416}; \citealt{Bonamigo_2018_RXCJ2248_M0416_M1206}; \citealt{Bergamini_2019_RXCJ2248_M0416_M1206}; \citealt{Caminha_2019_CLASHSLmass}; \citealt{Caminha_2023_ElGordoSL}; \citealt{Bergamini_2023_A2744SL}), from which we derive the cumulative total mass profiles to be studied. The main purpose of this paper is to investigate whether the total mass profiles of our sample of galaxy clusters can be described by one (or more) spherically symmetric parametric models. We also test if any of the models that better reproduce the mass profiles is related to a finite value of the total mass for $r\rightarrow + \infty$. When it occurs, we look for a relation between the total mass of the galaxy clusters obtained from our models and the respective values of $M_{200\mathrm{c}}$\footnote{$M_{200\mathrm{c}}$ corresponds to the total mass whithin a sphere inside which the mean mass density is 200 times the value of the critical mass density of the Universe at the redshift of the cluster.} derived from the weak lensing analysis (e.g. \citealt{Umetsu_2018_CLASHmassWL}).  

The paper is organized as follows. In Section 2 we present the galaxy clusters sample, the used dataset, and the derived total projected mass profiles from previously published strong lensing galaxy clusters studies. In Section 3 we present all the models employed to reconstruct the projected mass profiles and the methods we use to perform the numerical fits. In Section 4 we present the results of the numerical fits of the total mass profiles, and we discuss about the performance of our models. In Section 5 we present the relation that we find between the total masses from our models and the $M_{200\mathrm{c}}$ values from the literature. In Section 6, we discuss our results and the interplay between our models and the adopted hypoteses.

Throughout the paper, we assume a flat $\Lambda$CDM cosmological model, in which the Hubble constant value is $H_0 = 70 \ \mathrm{km \, s} ^{-1} \ \mathrm{Mpc}^{-1}$ and the total matter density value is $\Omega_\mathrm{m}$ = 0.3. 

\section{Data sample}

\begin{figure*}
\begin{subfigure}
\centering 
\includegraphics[scale=0.235]{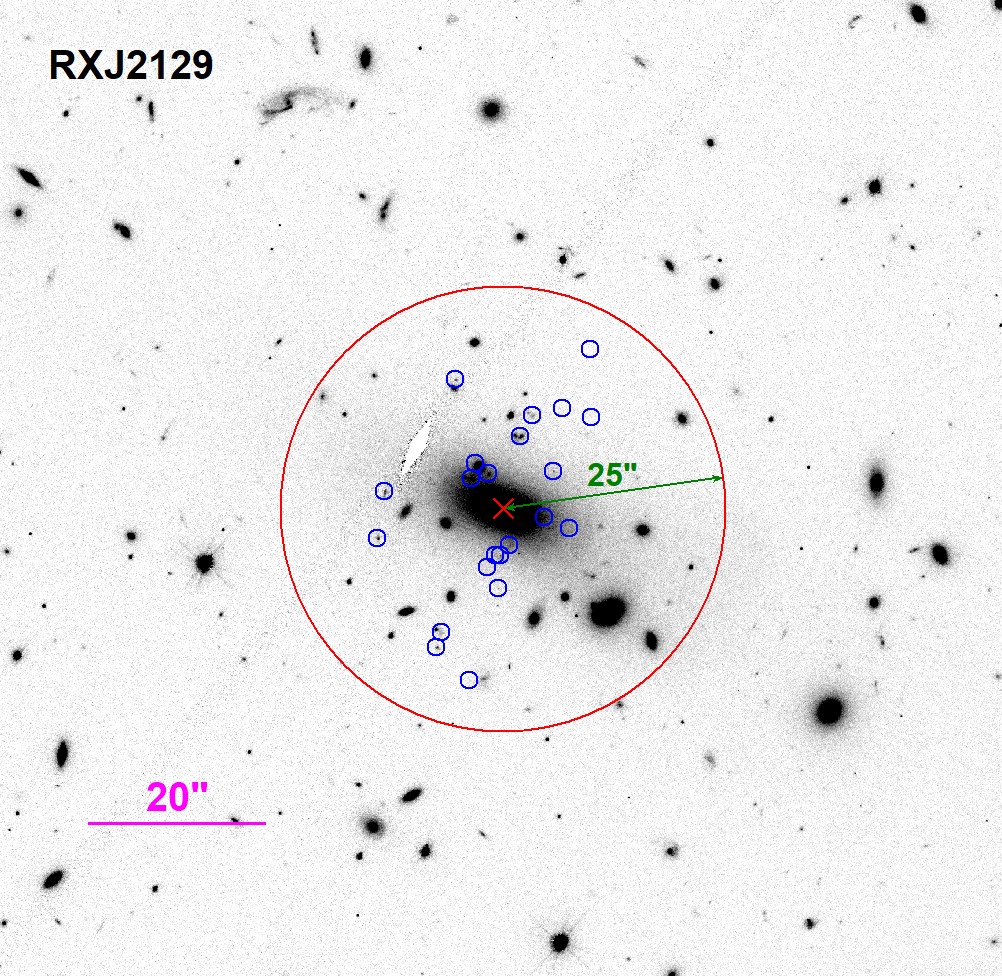}
\end{subfigure}
\begin{subfigure}
\centering 
\includegraphics[scale=0.22]{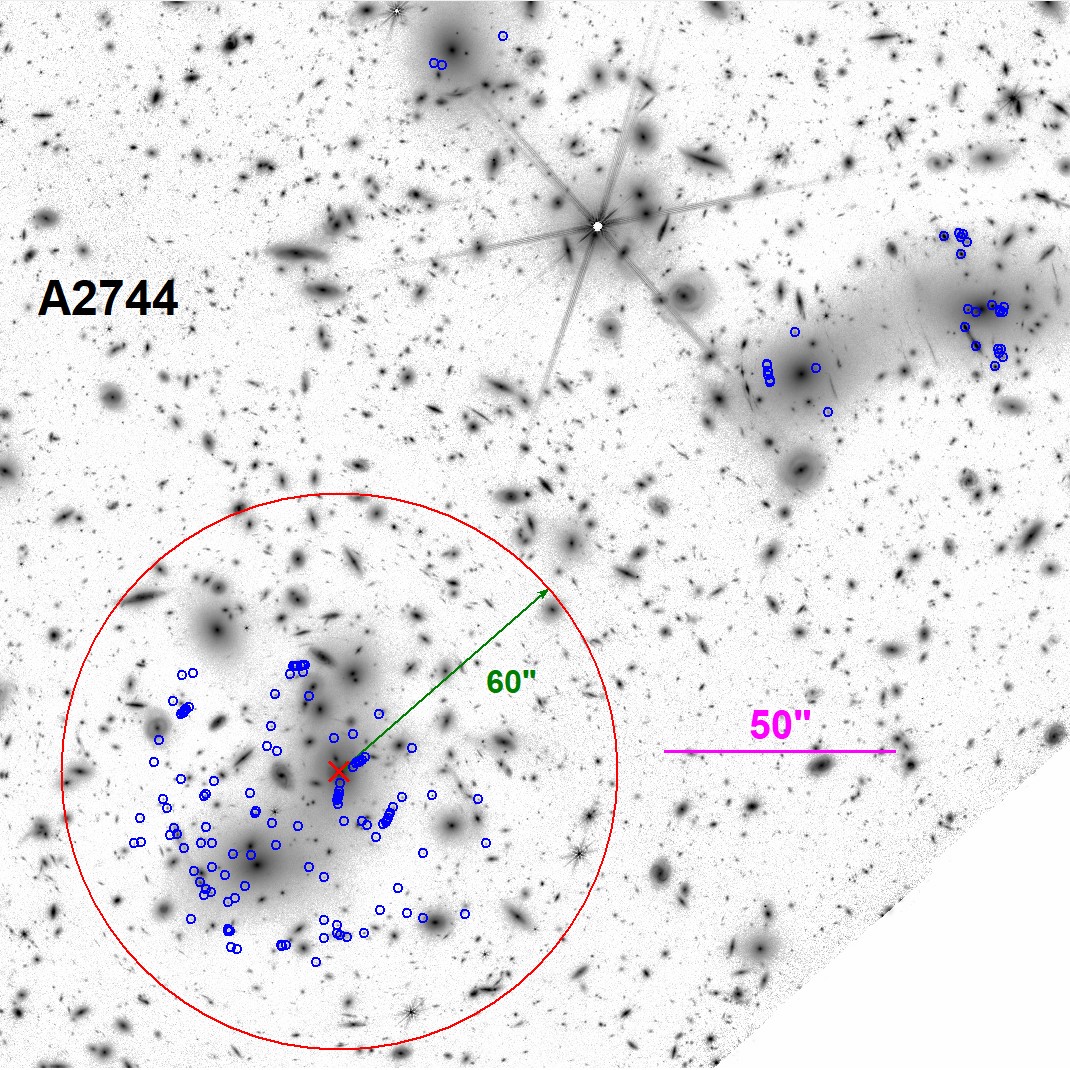}
\end{subfigure}
\begin{subfigure}
\centering 
\includegraphics[scale=0.22]{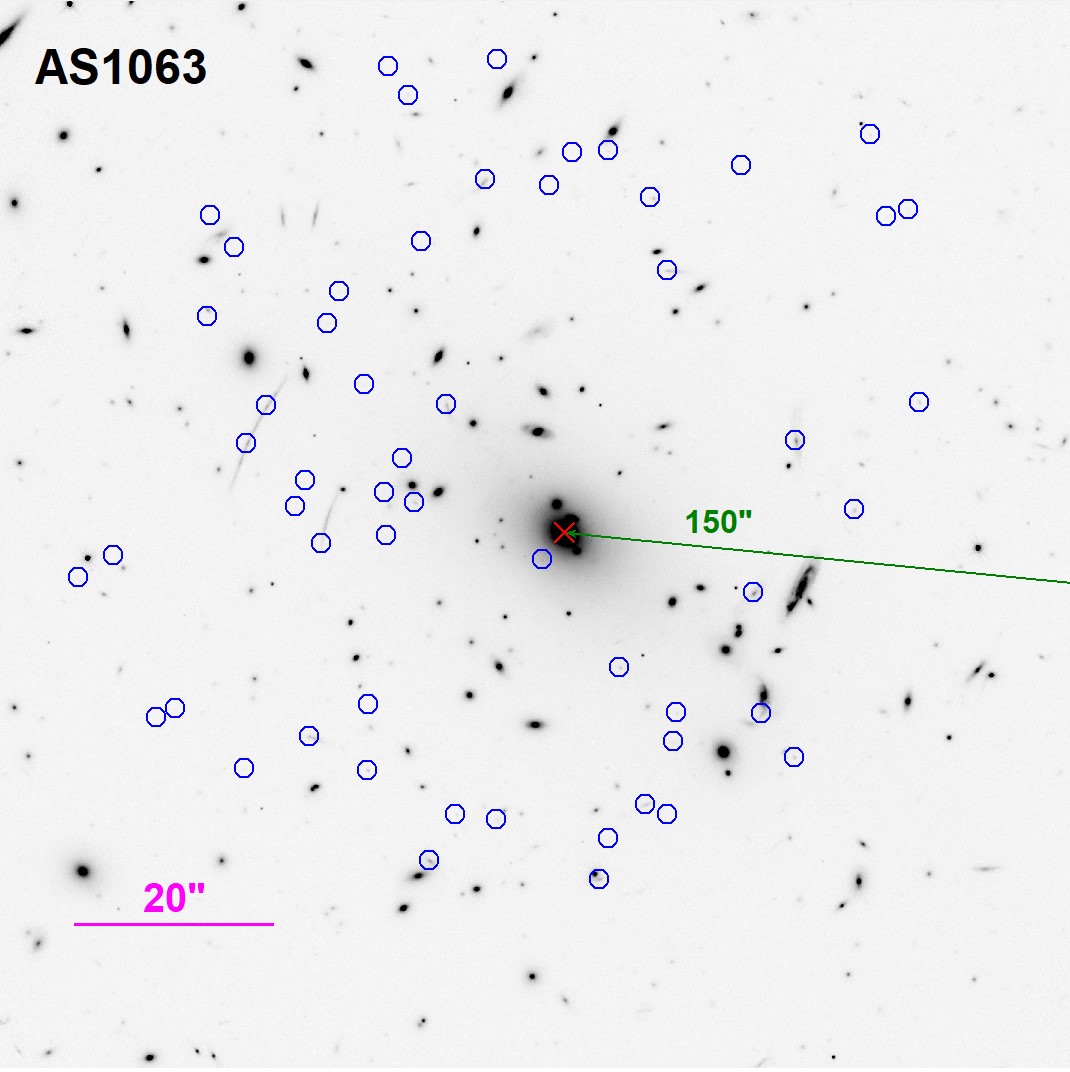}
\end{subfigure} \\
\begin{subfigure}
\centering 
\includegraphics[scale=0.22]{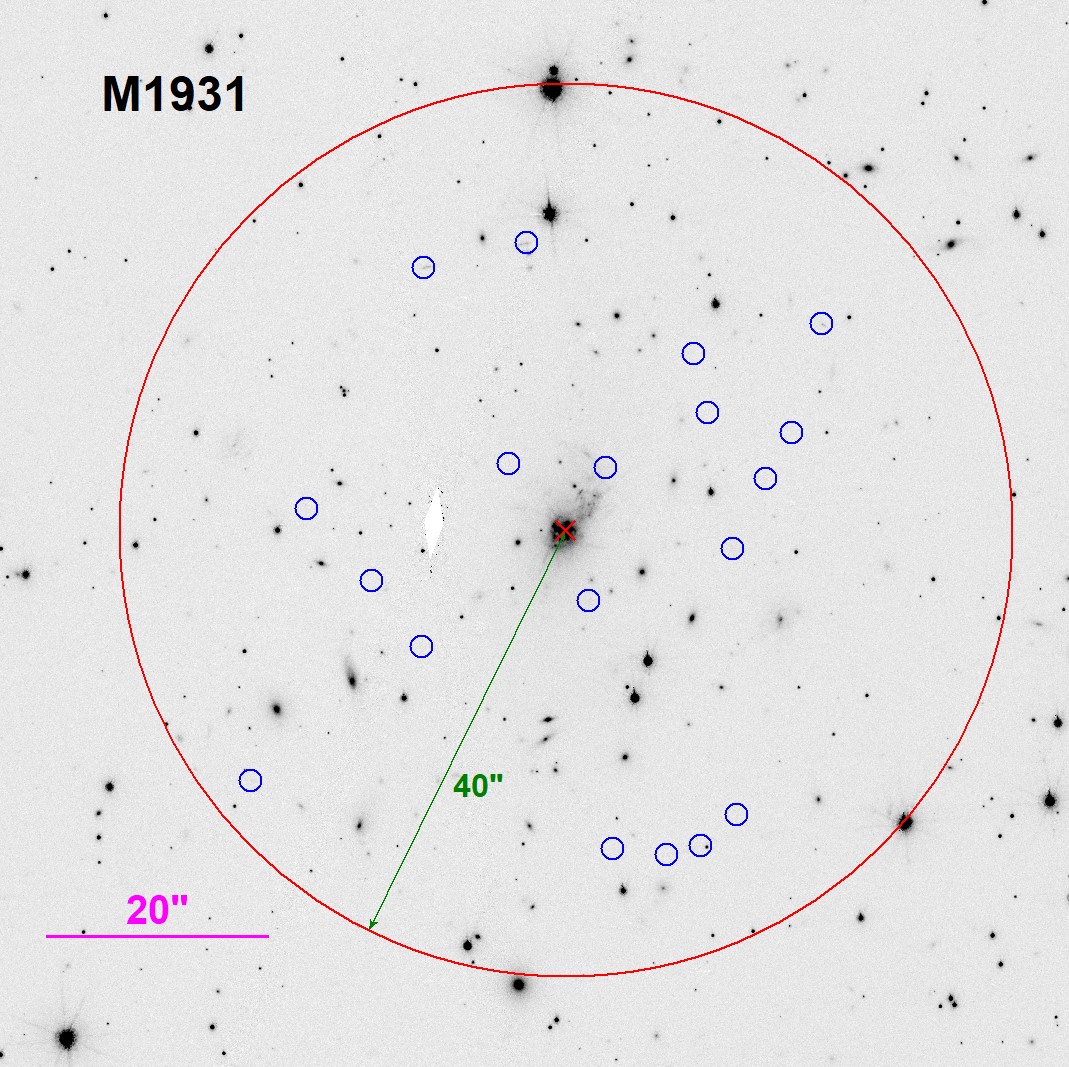}
\end{subfigure}
\begin{subfigure}
\centering 
\includegraphics[scale=0.22]{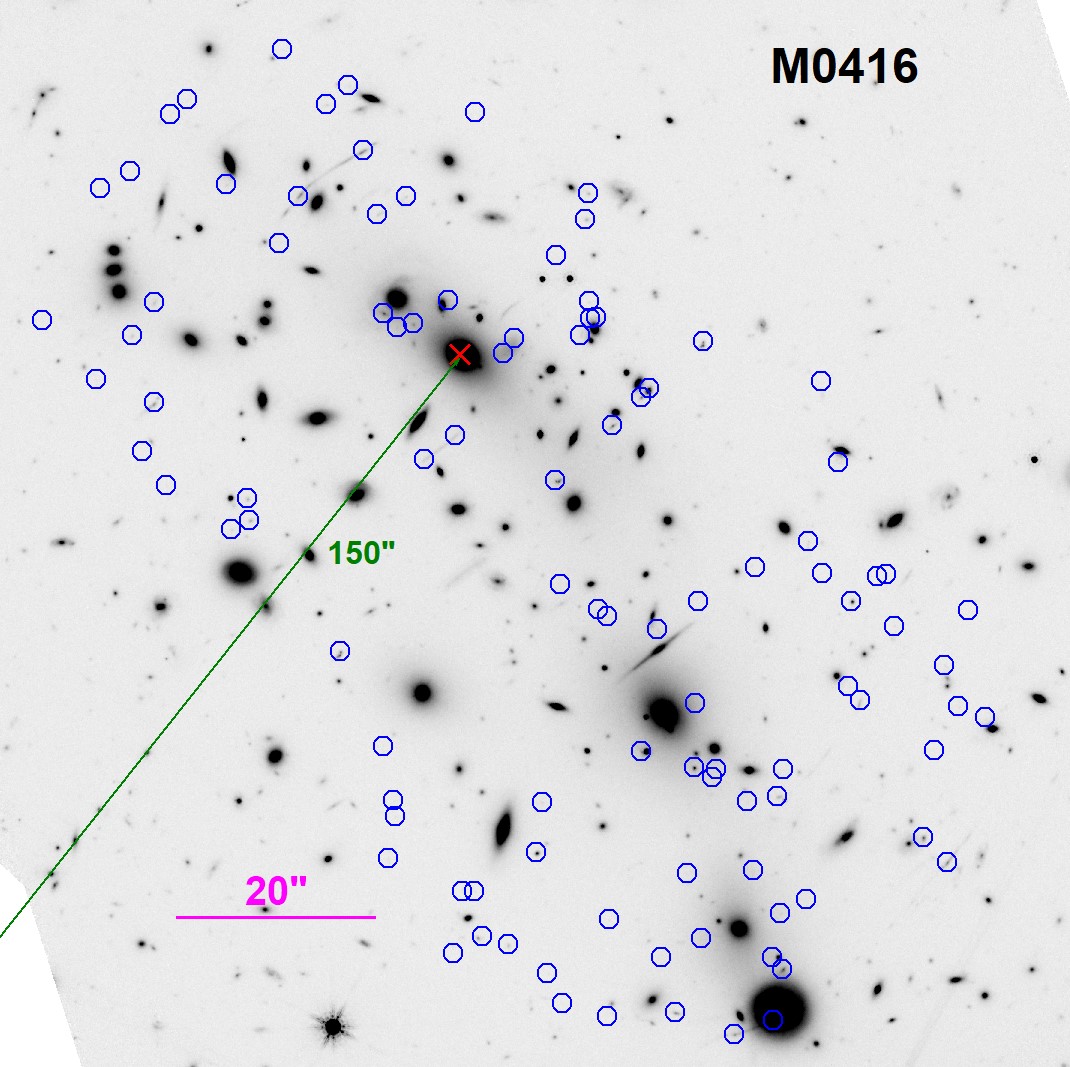}
\end{subfigure}
\begin{subfigure}
\centering 
\includegraphics[scale=0.22]{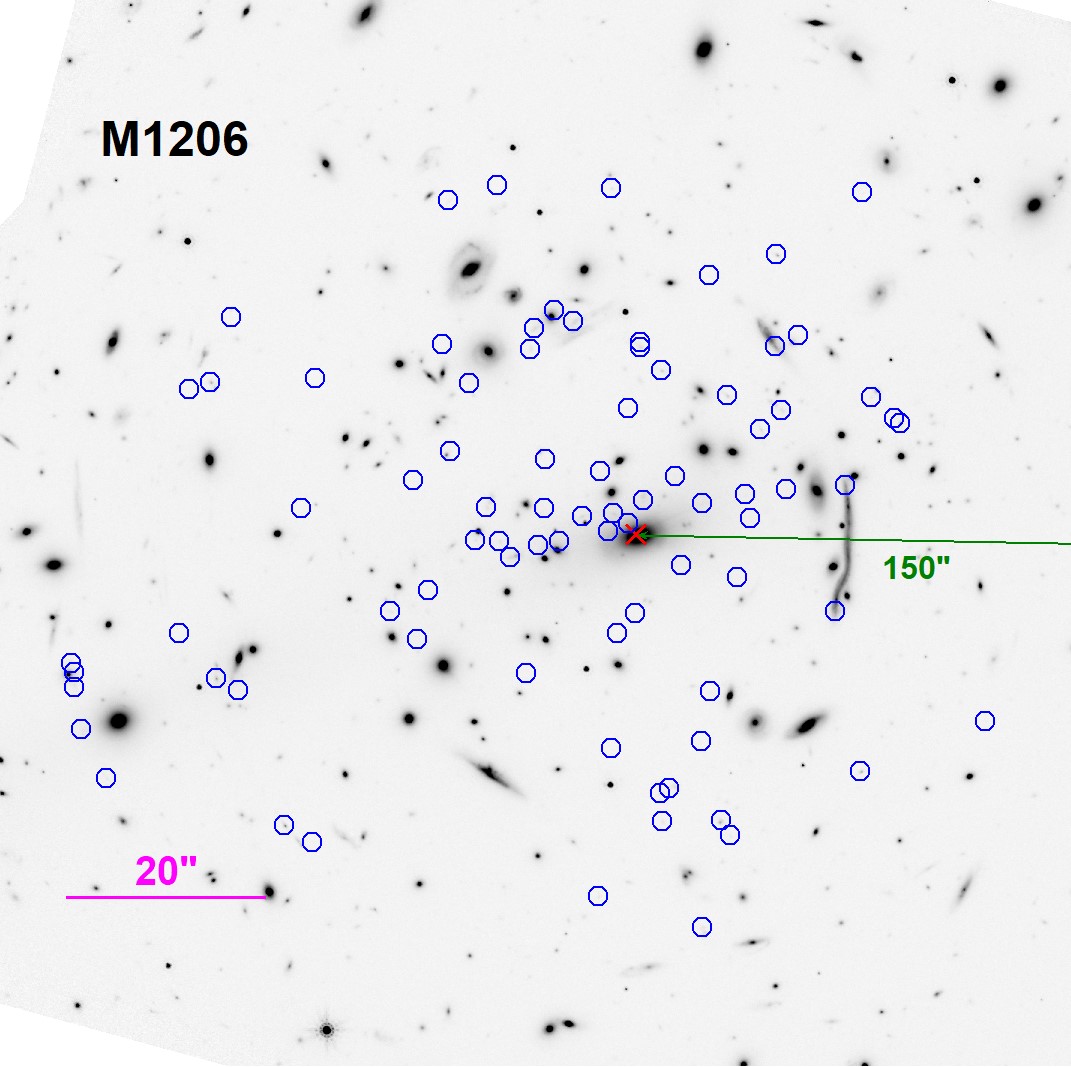}
\end{subfigure} \\
\begin{subfigure}
\centering 
\includegraphics[scale=0.22]{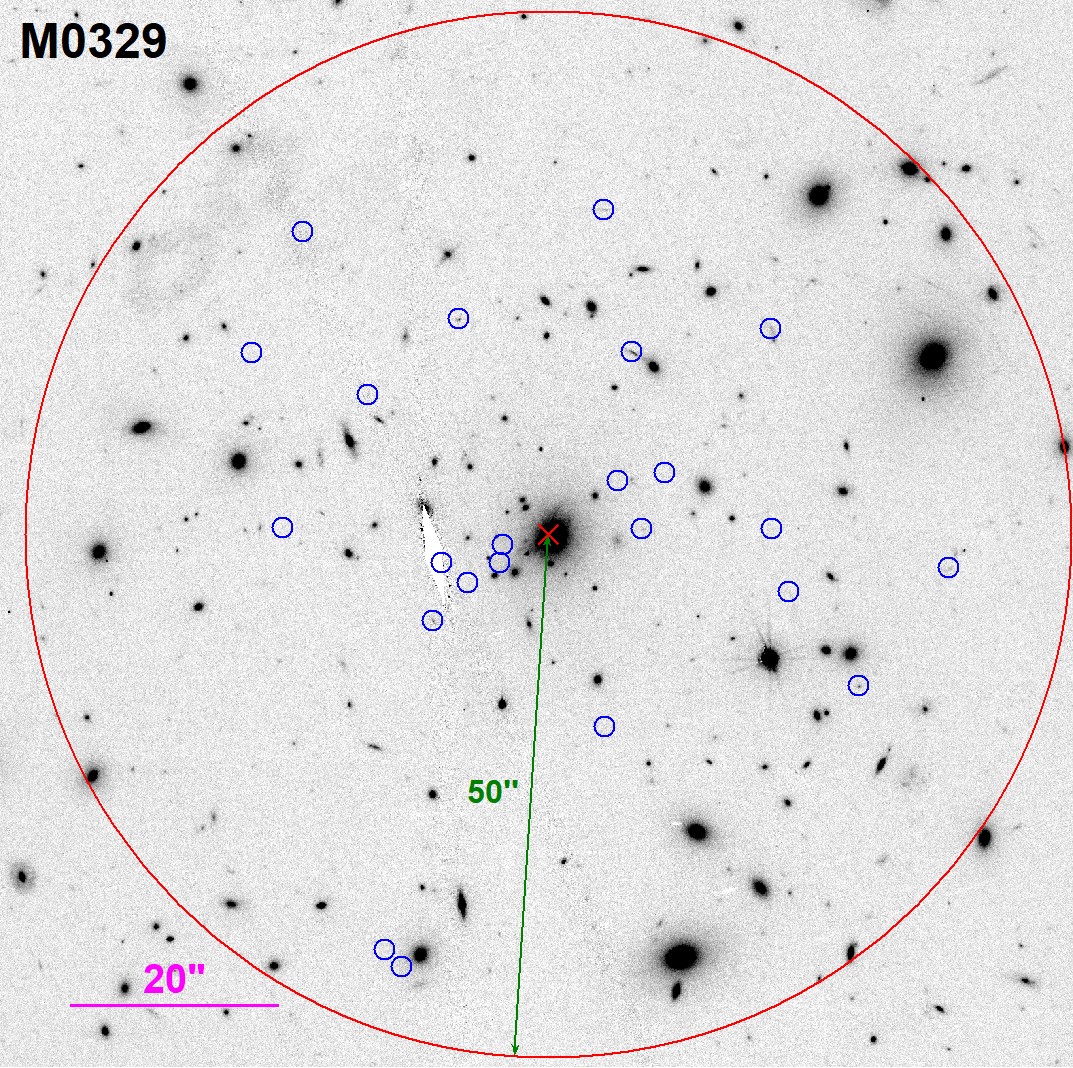}
\end{subfigure}
\begin{subfigure}
\centering 
\includegraphics[scale=0.22]{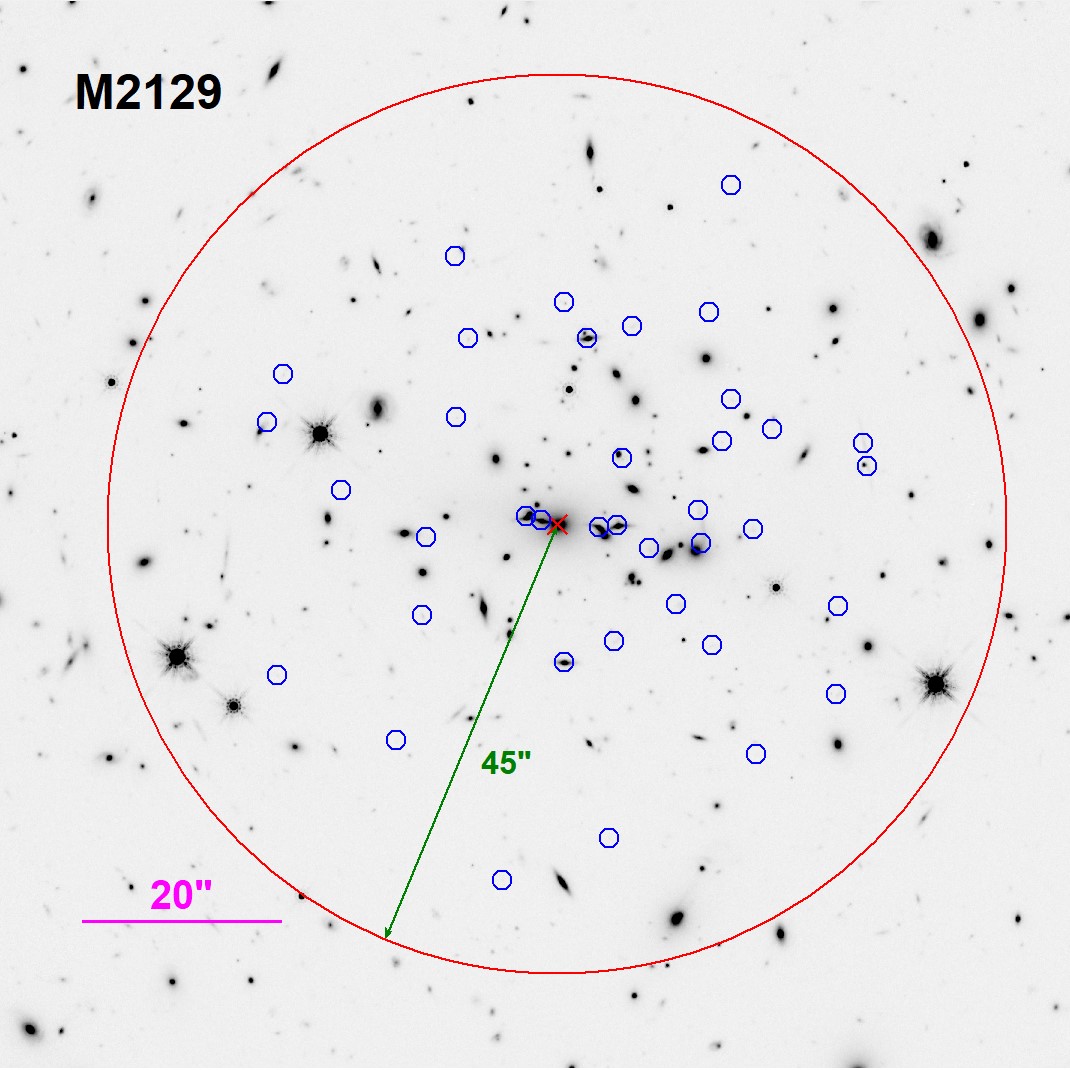}
\end{subfigure}
\begin{subfigure}
\centering 
\includegraphics[scale=0.22]{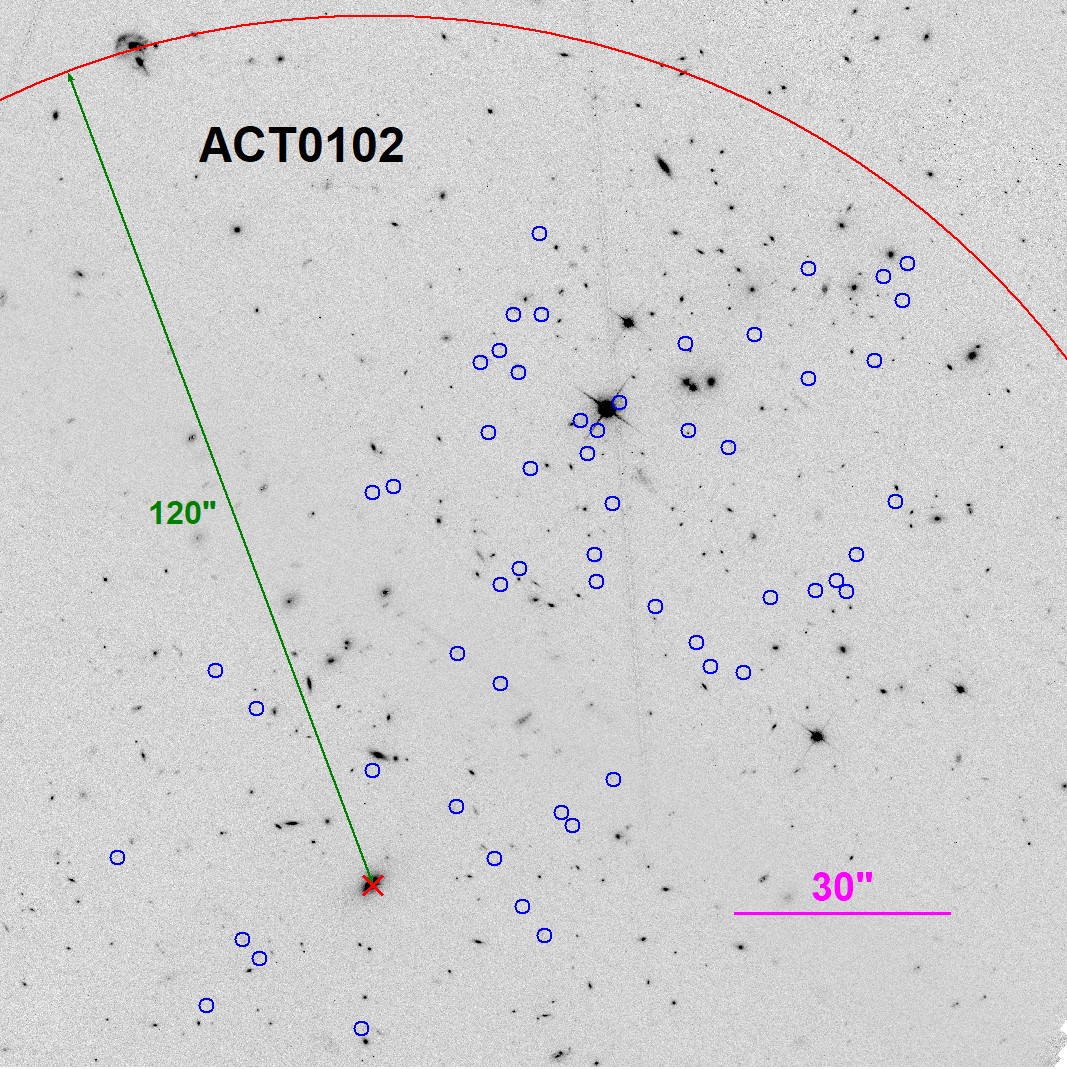}
\end{subfigure} \\
\caption{The nine galaxy clusters of our sample. The background images are from HST and JWST optical/infrared passbands. The blue circles represent the positions of the multiple images employed in the strong lensing studies by \cite{Caminha_2019_CLASHSLmass, Caminha_2023_ElGordoSL} and \cite{Bergamini_2019_RXCJ2248_M0416_M1206, Bergamini_2023_A2744SL}. The red circles, where visible, show the area surrounding the multiple images that we consider for the measurement of the projected total mass profiles. The red crosses indicate the centers of the red circles and the origin point of each projected total mass profile. The green lines represent the radii of the red circles, which length is shown near them.}
\label{fig.clusters}
\end{figure*}   

In this work, we study the nine galaxy clusters listed in Table \ref{tab.clusters}: RX J2129.7$+$0005 (namely RXJ2129), Abell 2744 (A2744), RXC J2248.7$-$4431 (AS1063), MACS J1931.8$-$2635 (M1931), MACS J0416.1$-$2403 (M0416), MACS J1206.2$-$0847 (M1206), MACS J0329.7$-$0211 (M0329), MACS J2129.4$-$0741 (M2129), and ACT-CL J0102$-$4915 (ACT0102 or El Gordo). These are massive galaxy clusters, with $M_{200\mathrm{c}}>5\times10^{14} M_\odot$, which span a wide redshift range ($0.2 < z < 0.9$). They are quite regular, virialized clusters observed by different photometric and spectroscopic observational campaigns, of which we report the details in the next sections. 

The nine galaxy clusters were selected to have high-quality and homogeneous data that allow for the development of accurate strong lensing models. The high-quality data have enabled the measurement of the cluster total mass values and cumulative mass profiles, which are employed herein.

\subsection{Photometric and spectroscopic data in SL models}

The cluster projected total mass measurements, as we show later, come from different studies that mainly rely on the observations made by the Cluster Lensing And Supernova survey with the HST (CLASH, \citealt{Postman_2012_CLASH}). All the clusters of our sample, except for A2744 and ACT0102, were targeted by CLASH, and this program is the main photometric source for them. In addition to CLASH, HST took images of M0416, A2744 and AS1063 within the Hubble Frontier Fields (HFF, \citealt{Lotz_2017_HFF}) program, significantly augmenting their photometric datasets by adding deeper exposures in seven filters (F435W, F606W, F814W, F105W, F125W,
F140W, F160W).

The photometry of A2744 is composite and it takes advantage of many different sources besides the HFF program, e.g. from the Beyond Ultra-deep Frontier Fields And Legacy Observations BUFFALO \citep{Steinhardt_2020_BUFFALO} program and from the recent observations of the GLASS-JWST \citep{Treu_2022_GLASSJWST} program. The complete list of sources of the photometry for A2744 can be found in Section 2.1 of \cite{Bergamini_2023_A2744SL}.

For ACT0102, the photometric observations employed in \cite{Caminha_2023_ElGordoSL} are provided by the Reionization Lensing Cluster Survey (RELICS, \citealt{Coe_2019_RELICS}) program.

On the spectroscopic side, the majority of data for the galaxy clusters in this paper is from the Multi Unit Spectroscopic Explorer (MUSE; \citealt{Bacon_2010_MUSEinstrument}) at the Very Large Telescope (VLT). MUSE observations are available for every galaxy cluster studied in this paper, and they have been exploited in the strong lensing analyses of \cite{Caminha_2019_CLASHSLmass, Caminha_2023_ElGordoSL} and \cite{Bergamini_2019_RXCJ2248_M0416_M1206, Bergamini_2023_A2744SL} (see these papers for a detailed description of each MUSE pointing). The MUSE exposure time for each galaxy cluster in the sample is at least of 2.45 hours, and this time was one of the determining factors for the choice of our galaxy cluster sample.

The other main spectroscopic data source for the galaxy clusters in this paper is the CLASH-VLT program \citep{Rosati_2014_CLASH}, that relied on the VIsual Multi-Object Spectrograph (VIMOS, \citealt{LeFevre_2003_VIMOS}). This program targeted all the galaxy clusters in our sample, except for A2744 and ACT0102.

Although A2744 is not part of CLASH-VLT, in addition to the data from MUSE its spectroscopic dataset comprehends other VIMOS observations \citep{Braglia_2009_A2744vimos}, the Grism Lens-Amplified Survey from Space (GLASS, \citealt{Treu_2015_GLASS}), and the AAOmega multi-object spectrograph \citep{Owers_2011_A2744mergers} (see \citealt{Bergamini_2023_A2744SL} for details). In \cite{Caminha_2023_ElGordoSL} instead, the strong lensing analysis of ACT0102 relies entirely on the MUSE observations.

\subsection{Strong lensing data} \label{sec.stronglensdata}

The cluster projected total mass profiles come from homogeneous strong lensing analyses in our sample. In particular, we extracted the cumulative total mass profiles from the mass models reconstructed by \cite{Caminha_2019_CLASHSLmass} for M0329, RXJ2129, M1931, M2129; \cite{Bergamini_2019_RXCJ2248_M0416_M1206} for AS1063, M0416, M1206; \cite{ Caminha_2023_ElGordoSL} for ACT0102, and \cite{Bergamini_2023_A2744SL} for A2744. In those works, the authors employed the robust photometric and spectroscopic information described above either to build new strong lensing models, like in \cite{Bergamini_2023_A2744SL}, or to enhance some existing models by adding new observations. As an example, in the analysis of AS1063, M0416, and M1206, \cite{Bergamini_2019_RXCJ2248_M0416_M1206} combined stellar kinematic information of cluster members with strong lensing models based on large samples of spectroscopically confirmed multiple images. There, the kinematic measurements provided specific priors for the free parameters of the lens model describing the total mass distribution of the cluster members. This additional effort allowed the authors to improve the previous models implemented in \cite{Caminha_2016_AS1063, Caminha_2017_M0416, Caminha_2017_M1206} for AS1063, M0416 and M1206, respectively, by reducing some degeneracies among the free parameters.

The total mass distribution of a cluster in parametric lens models is generally divided into a widespread component on cluster-scale, made of one or more DM halos, and a subhalo component on smaller scales tracing the cluster galaxies \citep{Natarajan_1997_lensingmodels}. Usually, the cluster-scale mass components are modeled with pseudo-isothermal elliptical mass distributions (PIEMD; \citealt{KassiolaKovner_1993_dPIE}), while the galaxy-scale subhalos are modeled with dual pseudo-isothermal mass density profiles (dPIE; \citealt{Eliasdottir_2007_dPIE}). In order to reduce the number of free parameters of the model, it is generally assumed that the parameters of the sub-halo population (the cluster member central velocity dispersion, $\sigma_\mathrm{v, gal}$, and truncation radius, $r_\mathrm{cut, gal}$) follow a scaling
relation that is compatible with the so-called tilt of the Fundamental Plane:
\begin{equation}
\label{eq.scalerel}
    \sigma_\mathrm{v, gal}=\sigma_\mathrm{v, gal}^\mathrm{ref} \left(\frac{L_\mathrm{gal}}{L_\mathrm{ref}}\right)^\alpha \, ; \, r_\mathrm{cut, gal}=r_\mathrm{cut, gal}^\mathrm{ref} \left(\frac{L_\mathrm{gal}}{L_\mathrm{ref}}\right)^\beta,
\end{equation}
where $L_\mathrm{gal}$ is the luminosity of the cluster member and $L_\mathrm{ref}$ is the reference luminosity; $\sigma_\mathrm{v, gal}^\mathrm{ref}$ and $r_\mathrm{cut, gal}^\mathrm{ref}$ are assumed as free parameters. The cluster total mass distribution of every galaxy cluster was modeled with the publicly available software \texttt{lenstool} \citep{Kneib_1996_AS1063lenstool, Jullo_2007_lenstool, Jullo_2010_lenstool}. The characteristics of the strong lensing model are different for each galaxy cluster. In Table \ref{tab.clusters}, we report some relevant information of the SL analysis, such as the number of cluster members included in each lensing model and the number of multiple images employed as observables. In the last column of Table \ref{tab.clusters}, we report the values of the total root-mean-square separation between the model-predicted and observed positions of the multiple images ($\Delta_\mathrm{RMS}$). We note that the median $\Delta_\mathrm{RMS}$ value of the model is $<0.8$ arcsec. This feature, in addition to demonstrating the high quality of the strong lensing studies on which our paper is based, is reflected by the very small errors on the projected total mass profiles. These small errors, consequently, allow the free parameters of our models to have small statistical uncertainties, as we show in Section \ref{sec.risultati}.

For each cluster, we derive the cumulative projected total mass profiles from the corresponding strong lensing model, by considering circular regions with increasing radii on the projected plane of the sky and measuring the total mass value enclosed within every circle. The clusters AS1063, M1206, RXJ2129, M1931, M0329, and M2129 have a single brightest central galaxy (BCG), hence the circular regions are centered on it. For A2744, we decide to center our profile on the northern BCG of the main clump (cfr. BCG-N in \citealt{Bergamini_2023_A2744SL}); similarly, for M0416 we center the circular regions on the northern BCG, and for ACT0102 on the southern BCG. In order to test the robustness of our results, for the galaxy clusters A2744, M0416, M0329 and ACT0102, we test if the best-fit parameters of the deprojected mass models change significantly when the center of the circular regions is moved to another point. We postpone the full description and the outcomes of such tests to Section \ref{sec.robtests}. In Figure \ref{fig.clusters}, the maximum circular regions and circle centers are marked in red. 

In conclusion, the final dataset for each cluster consists in $\sim$100 measurements of its projected cumulative total mass with increasing radial distance ($M_\mathrm{SL}(R)$ hereafter), accompanied by their intervals at the 68\%, 95\% and 99.8\% confidence levels.

\begin{table*}
\centering
\caption{List of the galaxy clusters studied in this paper, with relevant information for the strong lensing analyses.}
\label{tab.clusters}
\begin{tabular}{ccccccccc} 
\toprule
Cluster & $z$ & kpc/$\arcsec$ & $M_{200\mathrm{c}}$ ($10^{15} \, \mathrm{M}_\odot$) & $N_\mathrm{memb}$ & $z_\mathrm{memb}$ & $z_\mathrm{source}$ & $N_\mathrm{im}$ $(N_\mathrm{fam})$ & $\Delta_\mathrm{RMS}$ ($\arcsec$) \\
\midrule
RX J2129.7$+$0005 & 0.234 & 3.72 & $0.78 \pm 0.24$ & 70 & [0.217, 0.250] & [0.679, 3.427] & 22 (7) & 0.20\\
Abell 2744 &  0.307 & 4.53 & $2.06 \pm 0.42$ & 177 & [0.280, 0.340] & [1.026, 9.756] & 149 (50) & 0.43\\
RXC J2248.7$-$4431 & 0.348 & 4.92 & $1.98 \pm 0.60$ & 222 & [0.335, 0.362] & [0.730, 6.112] & 55 (20) & 0.55\\
MACS J1931.8$-$2635 & 0.352 & 4.96 & $1.16 \pm 0.28$ & 120 & [0.334, 0.370] & [1.178, 5.339] & 19 (7) & 0.38\\
MACS J0416.1$-$2403 & 0.396 &  5.34 & $1.14 \pm 0.27$ & 193 & [0.382, 0.410] & [0.940, 6.145] & 102 (37) & 0.61\\
MACS J1206.2$-$0847 & 0.439 &  5.68 & $1.51 \pm 0.32$ & 258 & [0.425, 0.453] & [1.012, 6.060] & 82 (27) & 0.46\\
MACS J0329.7$-$0211 & 0.450 & 5.76 & $1.27 \pm 0.22$ & 106 & [0.431, 0.470] & [1.313, 6.170] & 23 (9) & 0.24\\
MACS J2129.4$-$0741 & 0.587 & 6.62 & N/A & 138 & [0.566, 0.608] & [1.048, 6.846] & 38 (11) & 0.56\\
ACT-CL J0102$-$4915 & 0.870 & 7.71 & $2.76 \pm 0.51$ & 263 & [0.835, 0.907] & [2.189, 5.952] & 56 (23) & 0.75\\
\bottomrule 
\end{tabular}
\tablefoot{For each cluster, the columns show from the left: 1) extended name; 2) (mean) redshift; 3) conversion factor from arcsec to kpc at the cluster redshift; 4) $M_{200\mathrm{c}}$ values, if available; 5) number of cluster members included in the 
 strong lensing model; 6) redshift range of the cluster members included in the model; 7) redshift range of the multiply imaged sources; 8) number of multiple images ($N_\mathrm{im}$) and number of multiple image families $(N_\mathrm{fam})$; 9) total root-mean-square separation between model-predicted and observed multiple images ($\Delta_\mathrm{RMS}$). The $M_{200\mathrm{c}}$ values for AS1063, M0416, M1206, RXJ2129, M1931, M0329, and M2129 come from \cite{Umetsu_2018_CLASHmassWL}, while the value for A2744 is taken from \cite{Medezinski_2016_A2744WL}, and that for ACT0102 from \cite{Jee_2014_ElGordoWL}.}
\end{table*}

\section{Mass models}

The accurate photometric and spectroscopic multiband data listed in the previous section allowed for deep inspections of the mass structure for each galaxy cluster in the sample. The different mass diagnostics employed by previous works on these clusters, such as strong lensing, weak lensing, galaxy and hot gas dynamics (see e.g. \citealt{Biviano_2013_M1206}; \citealt{Bonamigo_2018_RXCJ2248_M0416_M1206}; \citealt{Caminha_2019_CLASHSLmass}; \citealt{Sartoris_2020_AS1063}; \citealt{Bergamini_2023_A2744SL}) showed unprecedent details of the total mass structure of these clusters, particularly in their central regions. In some specific cases (see e.g. \citealt{Bonamigo_2017_M0416, Bonamigo_2018_RXCJ2248_M0416_M1206}), it was shown that these structures are composed by a set of smaller substructures. However, we decide here to simplify and model all the galaxy clusters by assuming one-component, spherically-symmetric mass models. As we will discuss in Section \ref{sec.discussion}, the symmetry breaking introduced by the aforementioned substructures does not seem to affect significantly the $M_\mathrm{SL}(R)$ profiles and their fits.

\subsection{Models for the total mass distribution} \label{sec.massmodels}

We test several mass models to fit the $M_\mathrm{SL}(R)$ profiles: these are all spherical models with 2 or 3 free parameters. In the following, we present all the employed parametric mass density profiles, $\rho (r)$, together with their corresponding three-dimensional cumulative mass profiles, $M(r)$, which are linked through the relation

\begin{equation}
\label{eq.rhotom}
M(r)=\int_0^r 4\pi r'^2 \rho (r') \mathrm{d}r'.
\end{equation}

We indicate with $\rho_\mathrm{0}$ and $r_\mathrm{s}$ (or $r_\mathrm{C}$ and $r_\mathrm{T}$) the values of the characteristic scale mass density and radius. In detail, we consider:
\begin{itemize}
\item The Navarro-Frenk-White \citep{NFW_1997_DMprofile} profile,
\begin{multline}
\rho_\mathrm{NFW}(r)=\frac{\rho_\mathrm{0}}{\frac{r}{r_\mathrm{s}}\left(1+\frac{r}{r_\mathrm{s}}\right)^2} \, ; \\
M_\mathrm{NFW}(r) = 4\pi\rho_\mathrm{0}r_\mathrm{s}^3\left[\ln{\left(1+\frac{r}{r_\mathrm{s}}\right)}- \frac{r}{r+r_\mathrm{s}}\right].
\label{eq.nfw}
\end{multline}
\item The Non-singular Isothermal Sphere (NIS) profile, 
\begin{equation}
\rho_\mathrm{NIS}(r)=\frac{\rho_\mathrm{0}}{4\pi} \frac{1}{1+\frac{r^2}{r_\mathrm{s}^2}} \ ; \ M_\mathrm{NIS}(r)=\rho_\mathrm{0}r_\mathrm{s}^3 \left[\frac{r}{r_\mathrm{s}}-\mathrm{arctan}\left(\frac{r}{r_\mathrm{s}}\right)\right].
\label{eq.nis}
\end{equation}
\item The dual Pseudo Isothermal Ellipsiodal (dPIE) profile (\citealt{KassiolaKovner_1993_dPIE}; \citealt{Eliasdottir_2007_dPIE}),
\begin{multline}
\rho_\mathrm{dPIE}(r)=\frac{\rho_\mathrm{0}}{\left(1+\frac{r^2}{r_\mathrm{C}^2}\right)\left(1+\frac{r^2}{r_\mathrm{T}^2}\right)} \, ; \\
M_\mathrm{dPIE}(r)=4\pi \rho_\mathrm{0} r_\mathrm{C}^2 r_\mathrm{T}^2 \frac{r_\mathrm{C} \mathrm{arctan}\left(\frac{r}{r_\mathrm{C}}\right) - r_\mathrm{T} \mathrm{arctan}\left(\frac{r}{r_\mathrm{T}}\right)}{r_\mathrm{C}^2-r_\mathrm{T}^2}.
\label{eq.dpie}
\end{multline}
\item The Herquist profile \citep{Hernquist_1990_profile},
\begin{equation}
\rho_\mathrm{H}(r)=\frac{\rho_\mathrm{0}}{4\pi} \frac{1}{\frac{r}{r_\mathrm{s}}\left(1+\frac{r}{r_\mathrm{s}}\right)^3} \ ; \ M_\mathrm{H}(r)=\frac{\rho_\mathrm{0}r_\mathrm{s}^3}{2} \frac{r^2}{(r_\mathrm{s}+r)^2}.
\label{eq.her}
\end{equation}
\item The beta profile (e.g. \citealt{Ettori_2000_Xraybetamodel}),
\begin{equation}
\rho_\mathrm{\beta}(r)=\frac{\rho_\mathrm{0}}{4\pi} \frac{1}{\left(1+\frac{r^2}{r_\mathrm{s}^2}\right)^{3 \beta /2}} \ ; \ M_\mathrm{\beta}(r)=\frac{1}{3} \rho_\mathrm{0}r^3 \, _2F_1 \left[\frac{3}{2},\frac{3\beta}{2}, \frac{5}{2}, -\frac{r^2}{r_\mathrm{s}^2}\right],
\label{eq.beta}
\end{equation}
where $_2F_1$ is the hypergeometric function \citep{Gauss_hypergeom} and $\beta$ is another free parameter in addition to $\rho_\mathrm{0}$ and $r_\mathrm{s}$.
\end{itemize}

\subsection{Projection and MCMC fitting} \label{sec.projMCMC}

To fit the cluster cumulative, projected total mass profiles, we adopt the following formula for the cumulative two-dimensional mass, $M(R)$, that is derived from the projection of the total mass density profile:

\begin{multline}
\label{eq.proj}
M(R)=\int_0^R 2 \pi R' \mathrm{d}R' \times 2\int_{R'}^{+\infty} \frac{\rho(r)r}{\sqrt{r^2-R'^2}} \mathrm{d}r = \\
=\int_0^R R' \mathrm{d}R' \int_{R'}^{+\infty} \frac{\mathrm{d}r}{r\sqrt{r^2-R'^2}} \frac{\mathrm{d}M(r)}{\mathrm{d}r},
\end{multline}
where $M(r)$ is the three-dimensional mass profile (see Equation \ref{eq.rhotom}) and

\begin{equation}
\label{eq.projrho}
2 \int_R^{+\infty} \frac{\rho (r) r \mathrm{d}r}{\sqrt{r^2-R^2}}=\Sigma (R)
\end{equation}
is the surface mass density. We note that we indicate with $R$ and $r$ the projected, two-dimensional radius and the three-dimensional radius, respectively.

In order to perform the numerical fits of the $M_\mathrm{SL}(R)$ profiles, we adopt a bayesian approach for the optimization of the free parameters. Specifically, we use the publicly available python library \texttt{emcee} \citep{emcee}, to perform the sampling of the posterior distributions of the model free parameters. The values presented as results in the next section correspond to the 50th percentile (the median) and to the 16th and 84th percentiles (the 1$\sigma$ confidence level errors) of the posterior distributions. Henceforth, we report the median values of the free parameters for each total mass model, and we consequently plot the corresponding profiles.

\section{Results} \label{sec.risultati}

\begin{figure*}
\begin{subfigure}
\centering 
\includegraphics[scale=0.24]{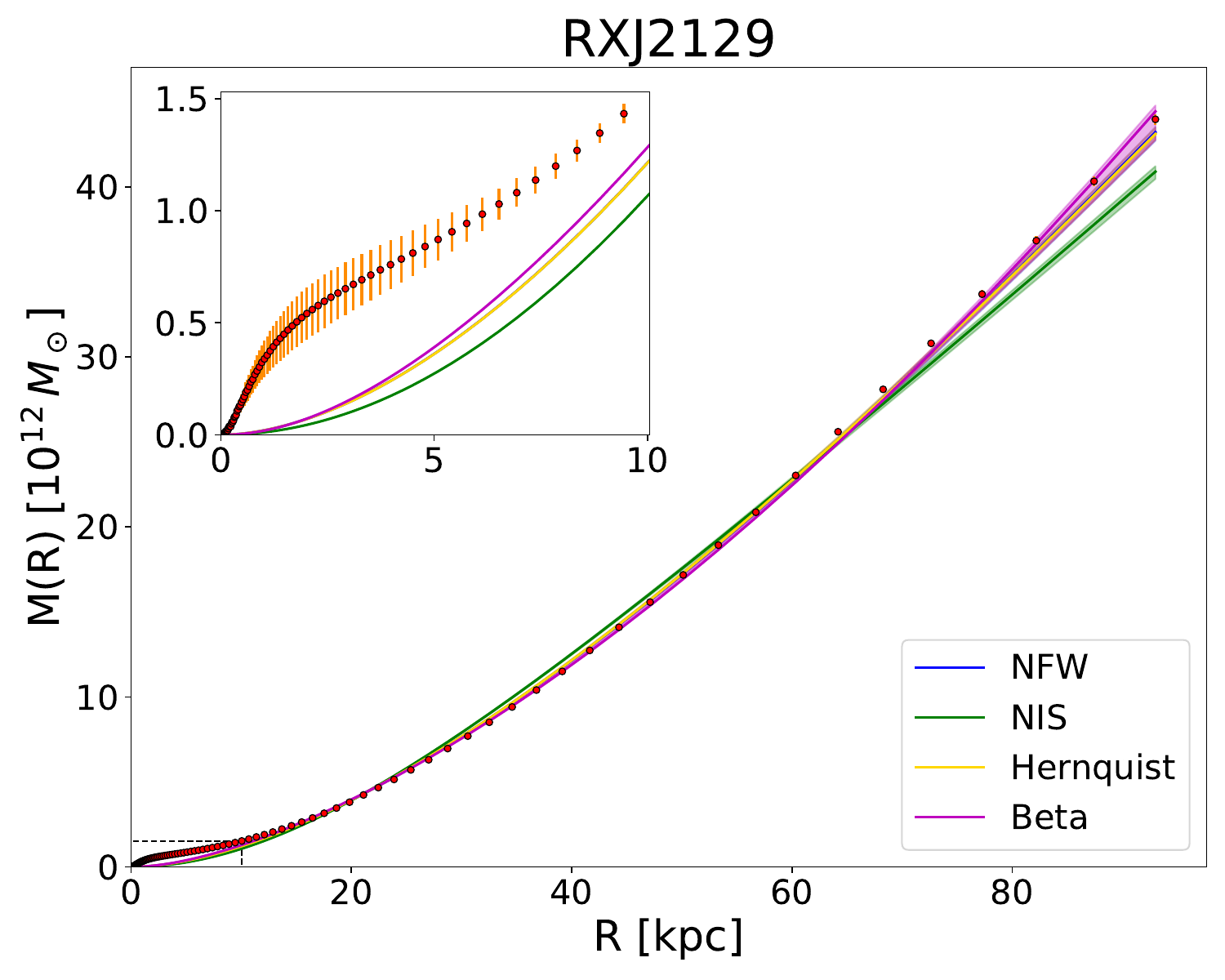}
\end{subfigure}
\begin{subfigure}
\centering 
\includegraphics[scale=0.24]{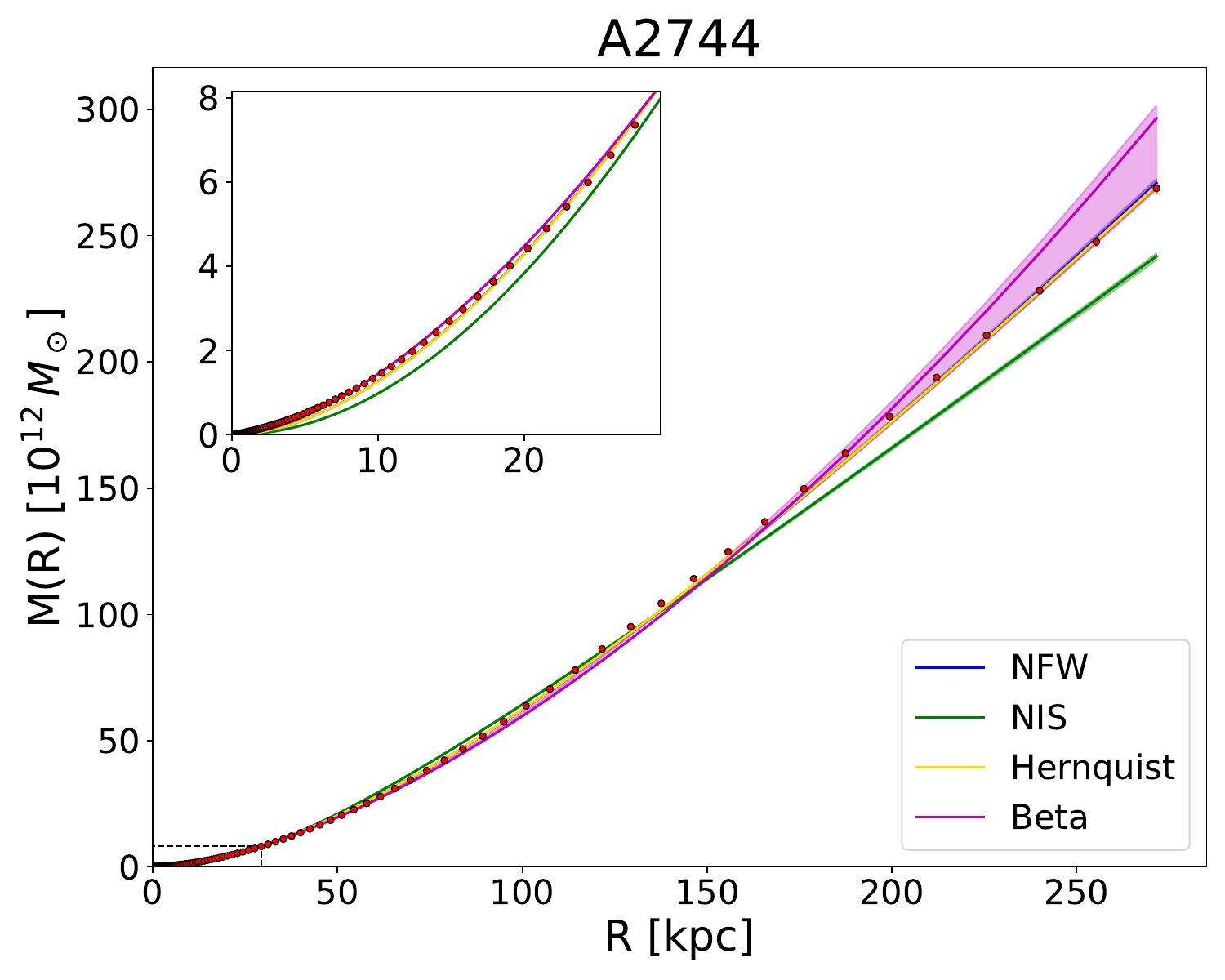}
\end{subfigure}
\begin{subfigure}
\centering 
\includegraphics[scale=0.24]{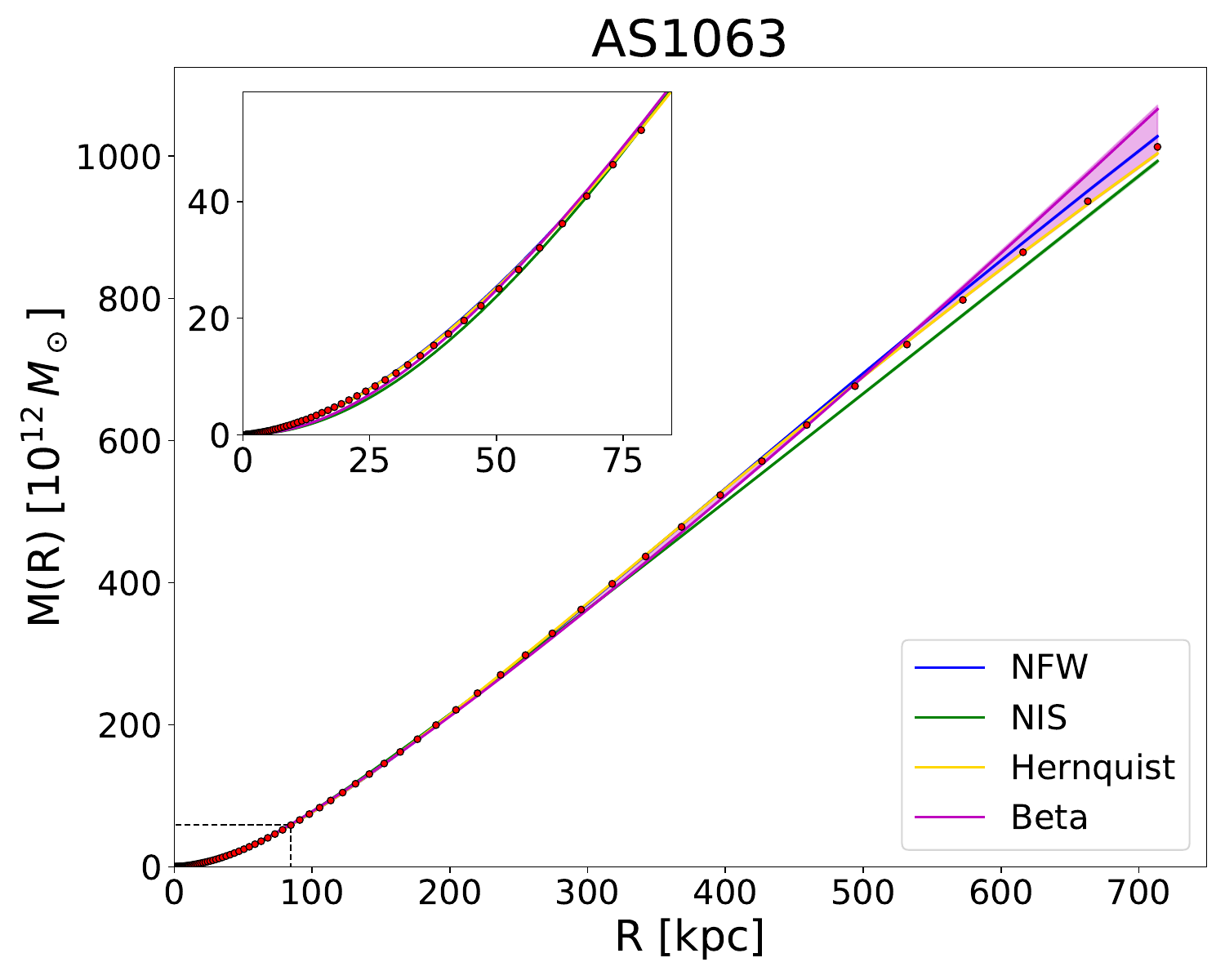}
\end{subfigure} \\
\begin{subfigure}
\centering 
\includegraphics[scale=0.24]{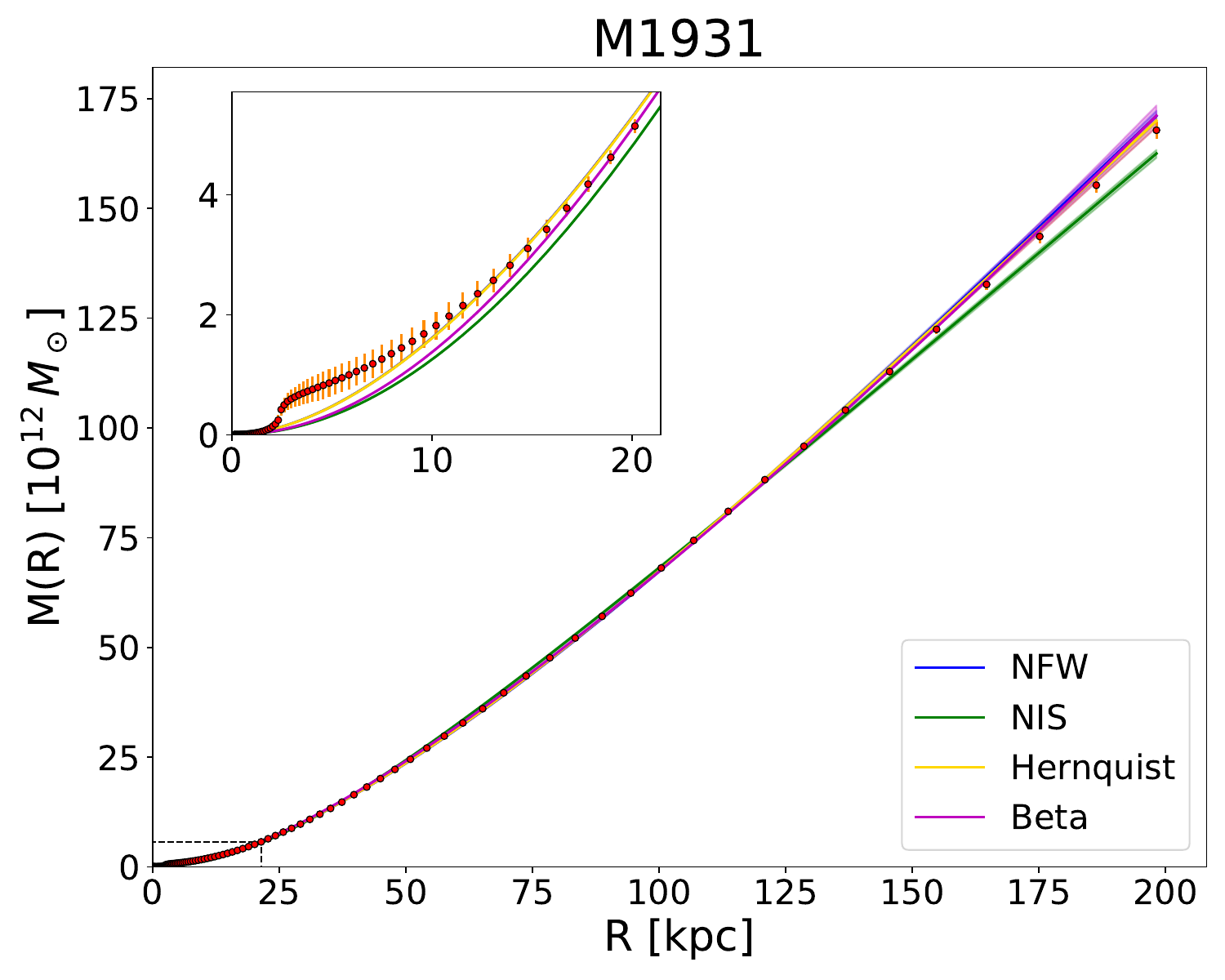}
\end{subfigure}
\begin{subfigure}
\centering 
\includegraphics[scale=0.24]{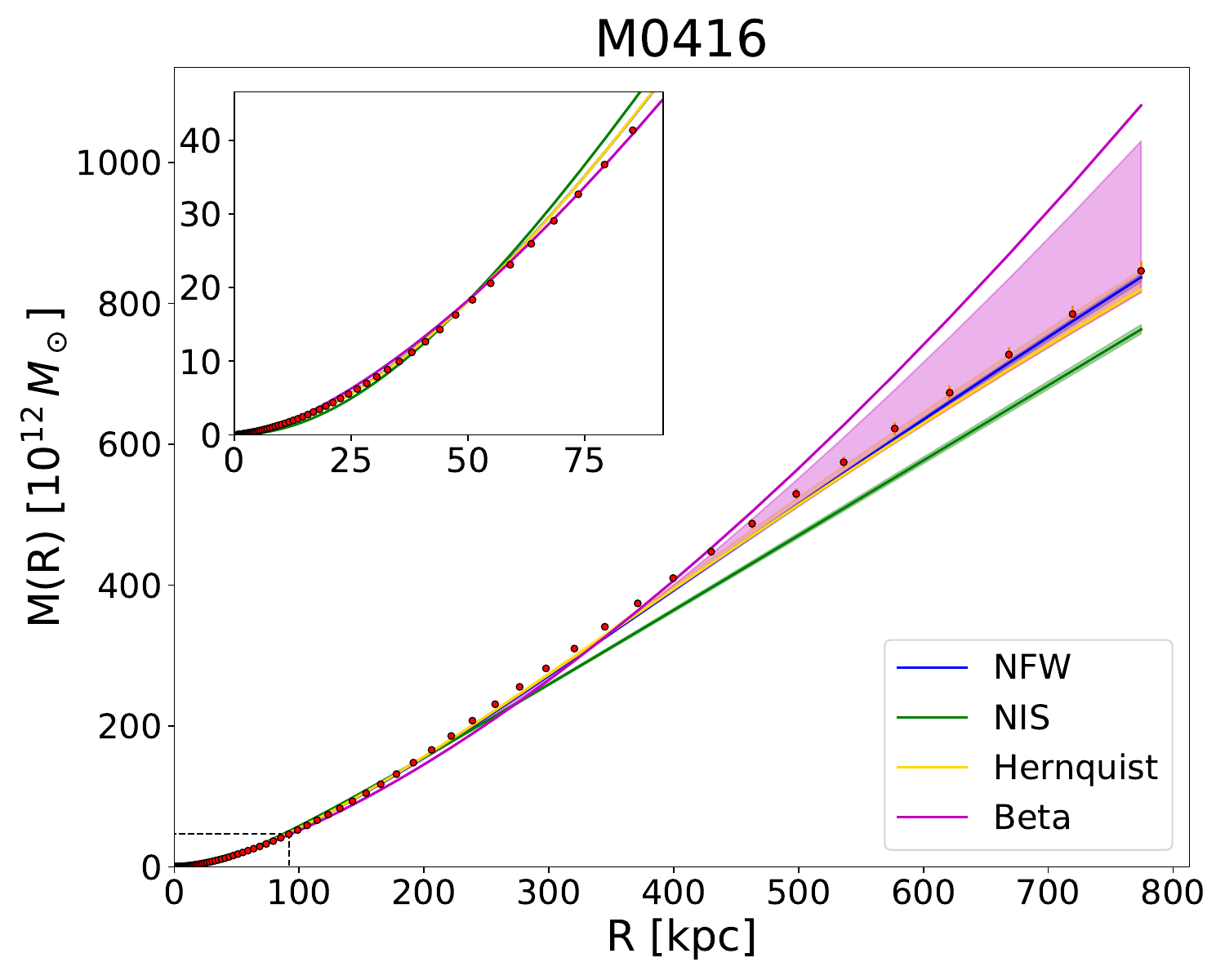}
\end{subfigure}
\begin{subfigure}
\centering 
\includegraphics[scale=0.24]{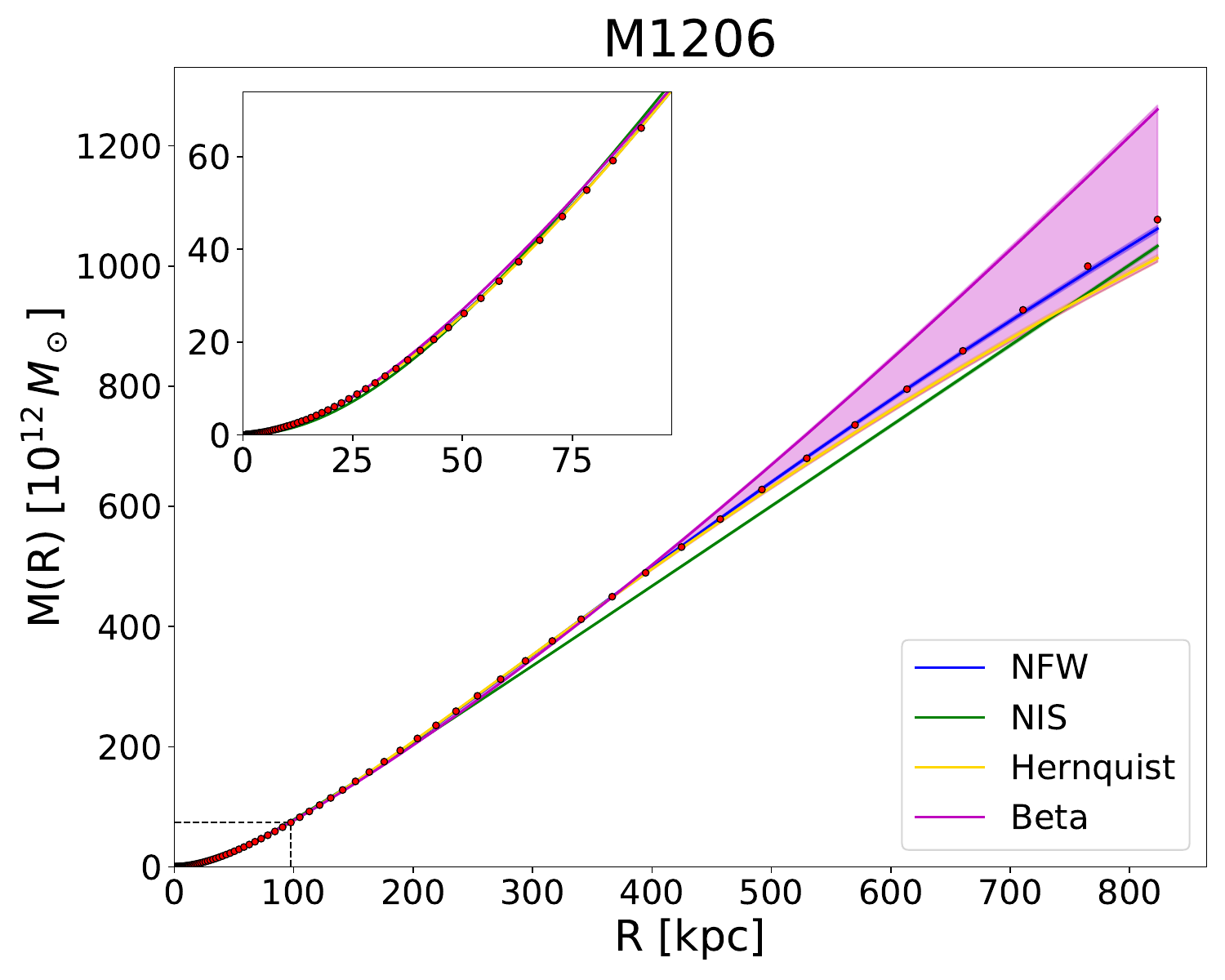}
\end{subfigure} \\
\begin{subfigure}
\centering 
\includegraphics[scale=0.24]{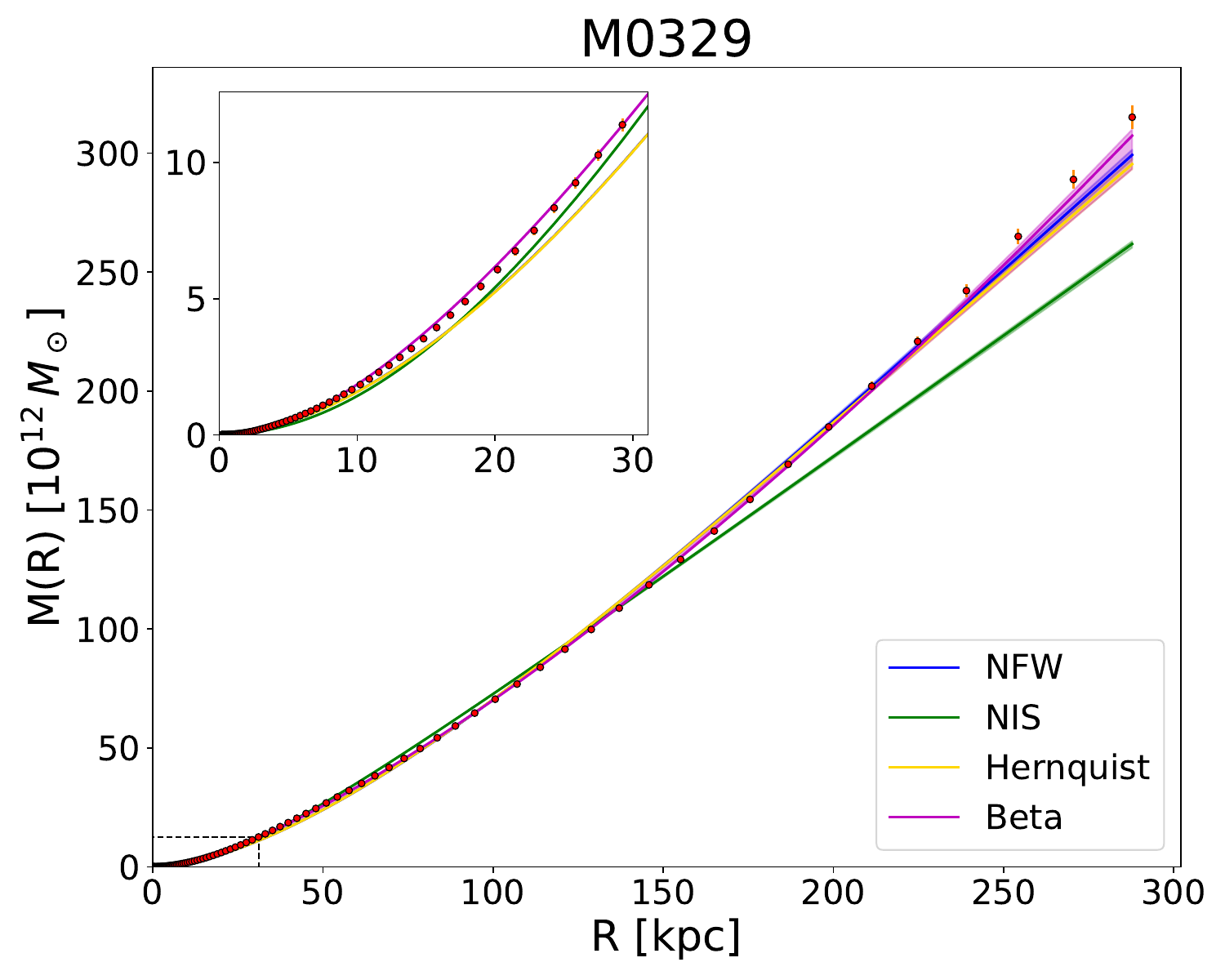}
\end{subfigure}
\begin{subfigure}
\centering 
\includegraphics[scale=0.24]{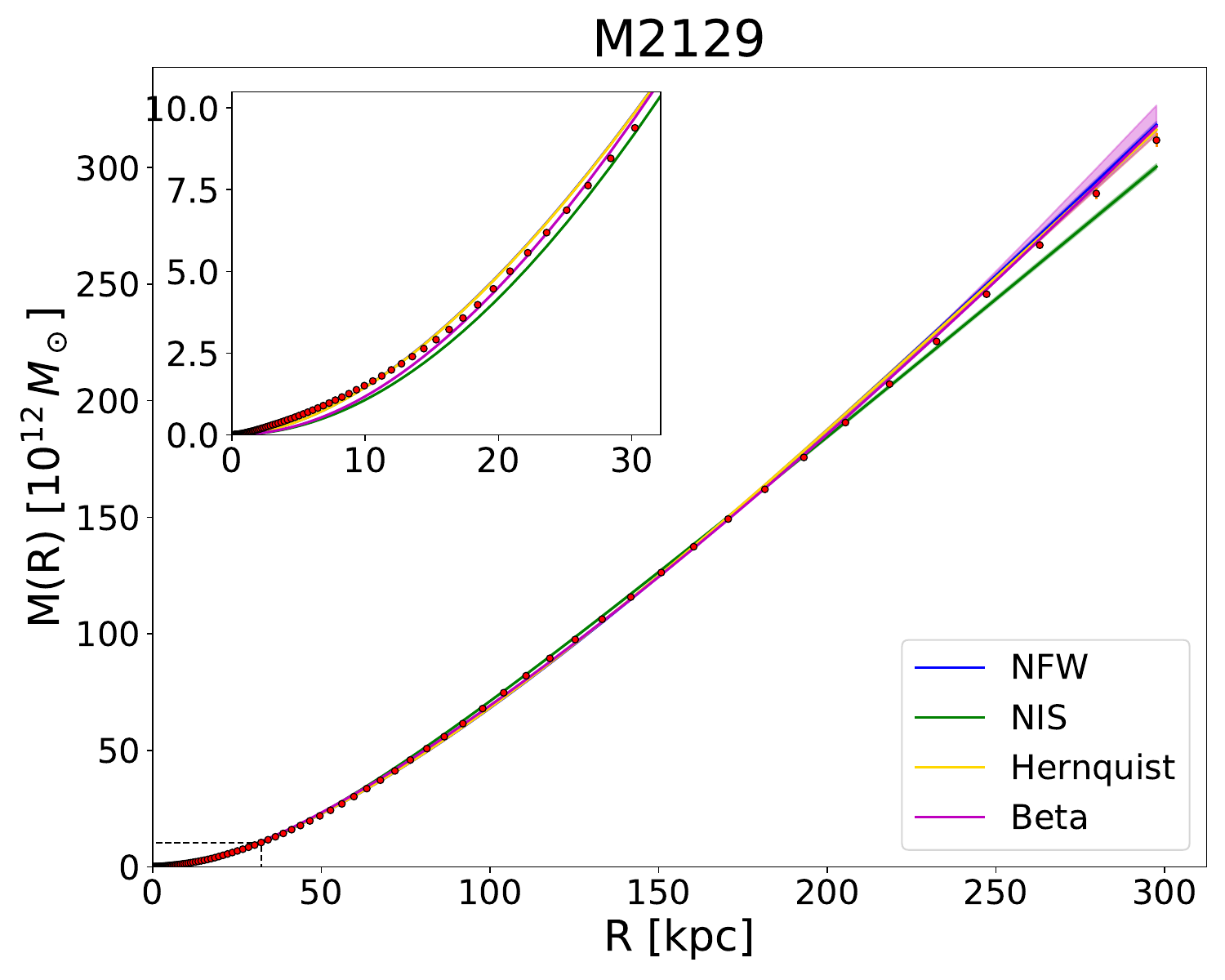}
\end{subfigure}
\begin{subfigure}
\centering 
\includegraphics[scale=0.24]{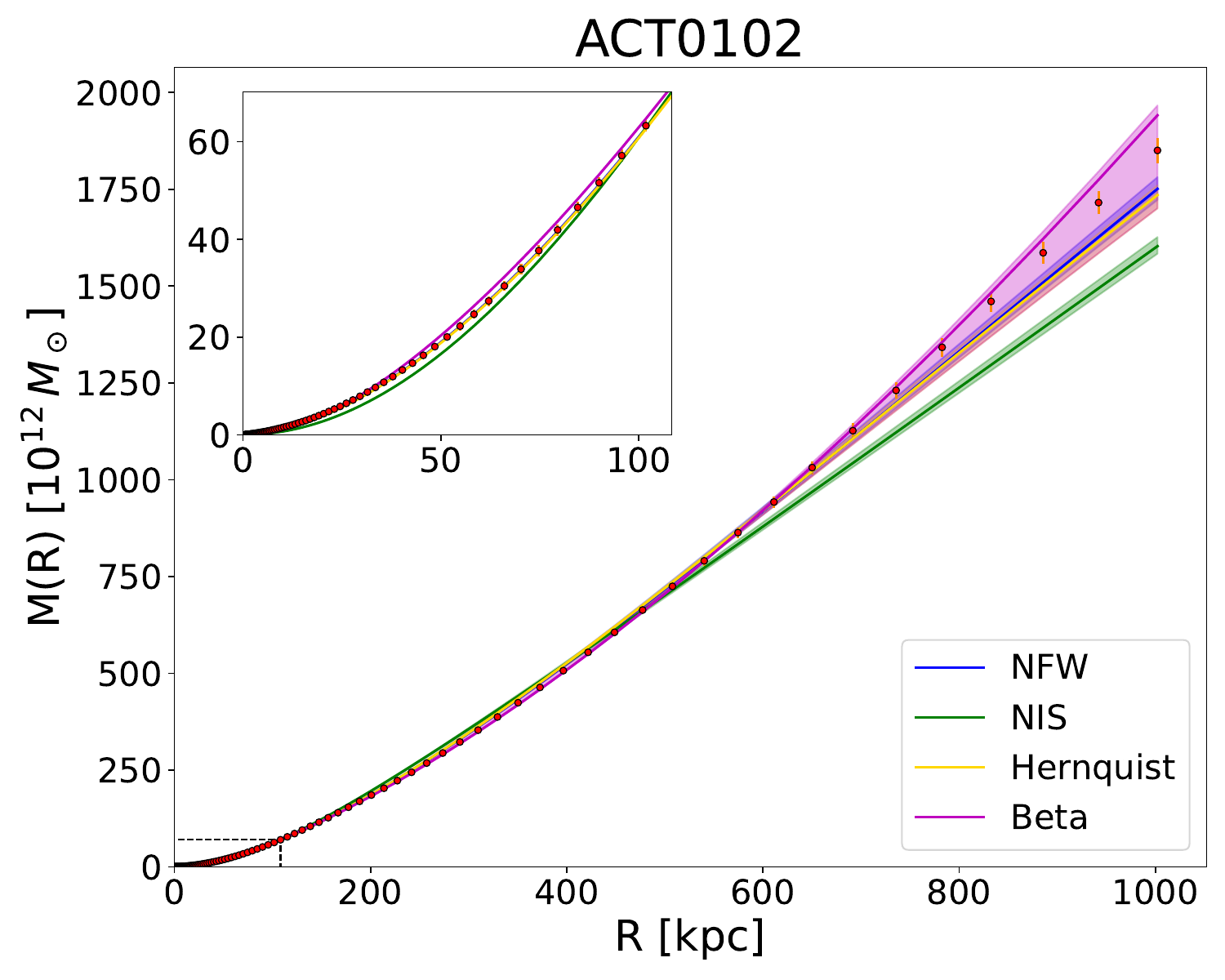}
\end{subfigure} \\
\caption{Cumulative projected total mass profiles for the nine clusters in the sample. The red dots are the measured values with their corresponding errors (1$\sigma$ error bars), while the solid lines represent the best-fits of the $M_\mathrm{SL}(R)$ profiles of each mass model (NFW, NIS, Hernquist, Beta). The shaded regions represent the 1$\sigma$ (68\%) confidence regions. Each dashed rectangle is zoomed in the upper-left corner of each plot.}
\label{fig.totprof}
\end{figure*} 

In Figure \ref{fig.totprof}, we show the cumulative projected total mass profiles for each cluster in the sample and the corresponding errors, together with the best-fitting profiles for each model and their respective uncertainties. The reported error bars and shadowed regions are computed with a two-step method. The first step consists in randomly selecting 100 sets of parameter values from the Monte Carlo Markov chains, and to generate with them the corresponding total mass profiles. The second step is to extract, at each projected radius $R$, the 16th and the 84th percentiles of the previously generated 100 total mass values, and to identify them as the confidence boundaries.

\begin{figure*}
\begin{subfigure}
\centering 
\includegraphics[scale=0.24]{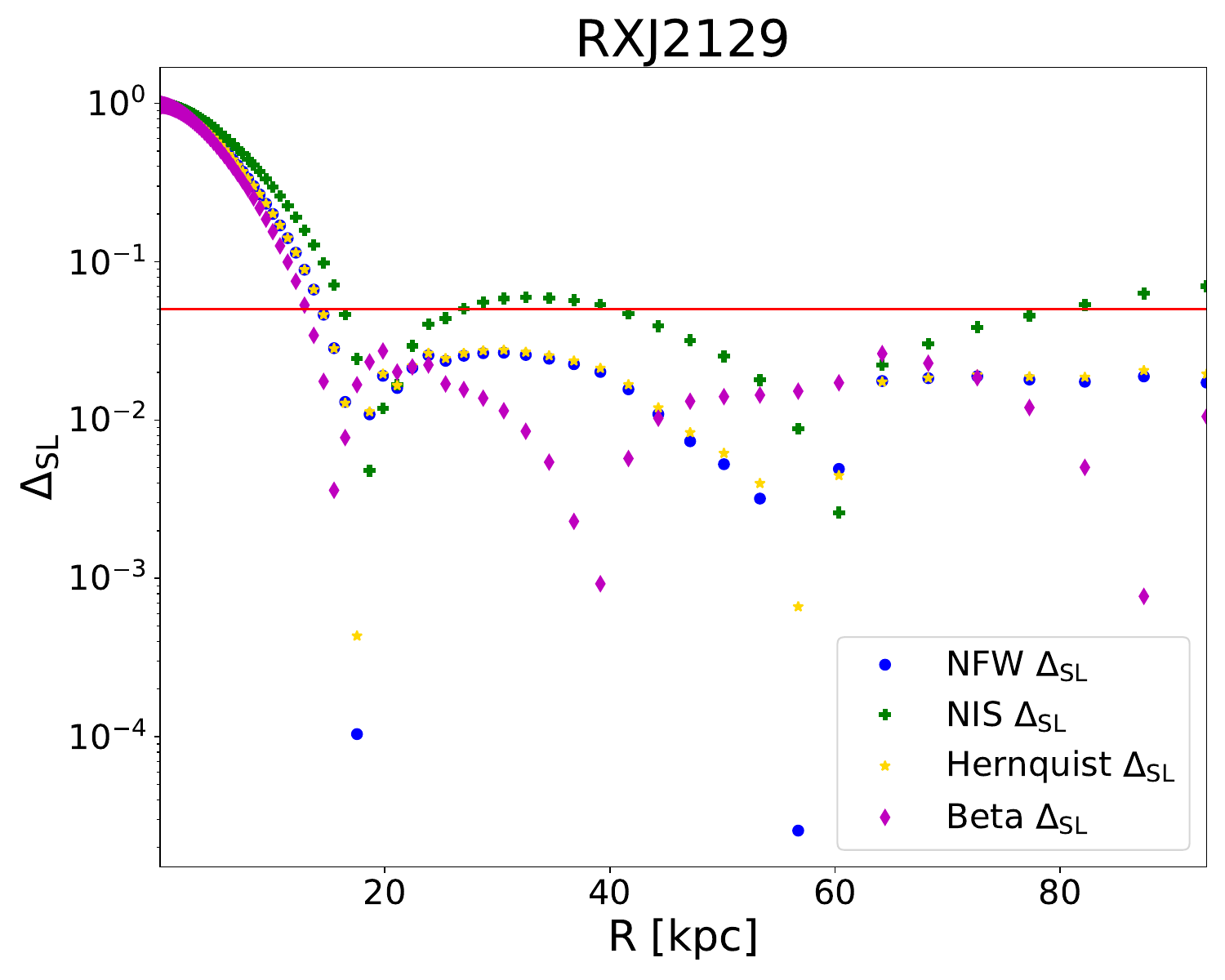}
\end{subfigure}
\begin{subfigure}
\centering 
\includegraphics[scale=0.24]{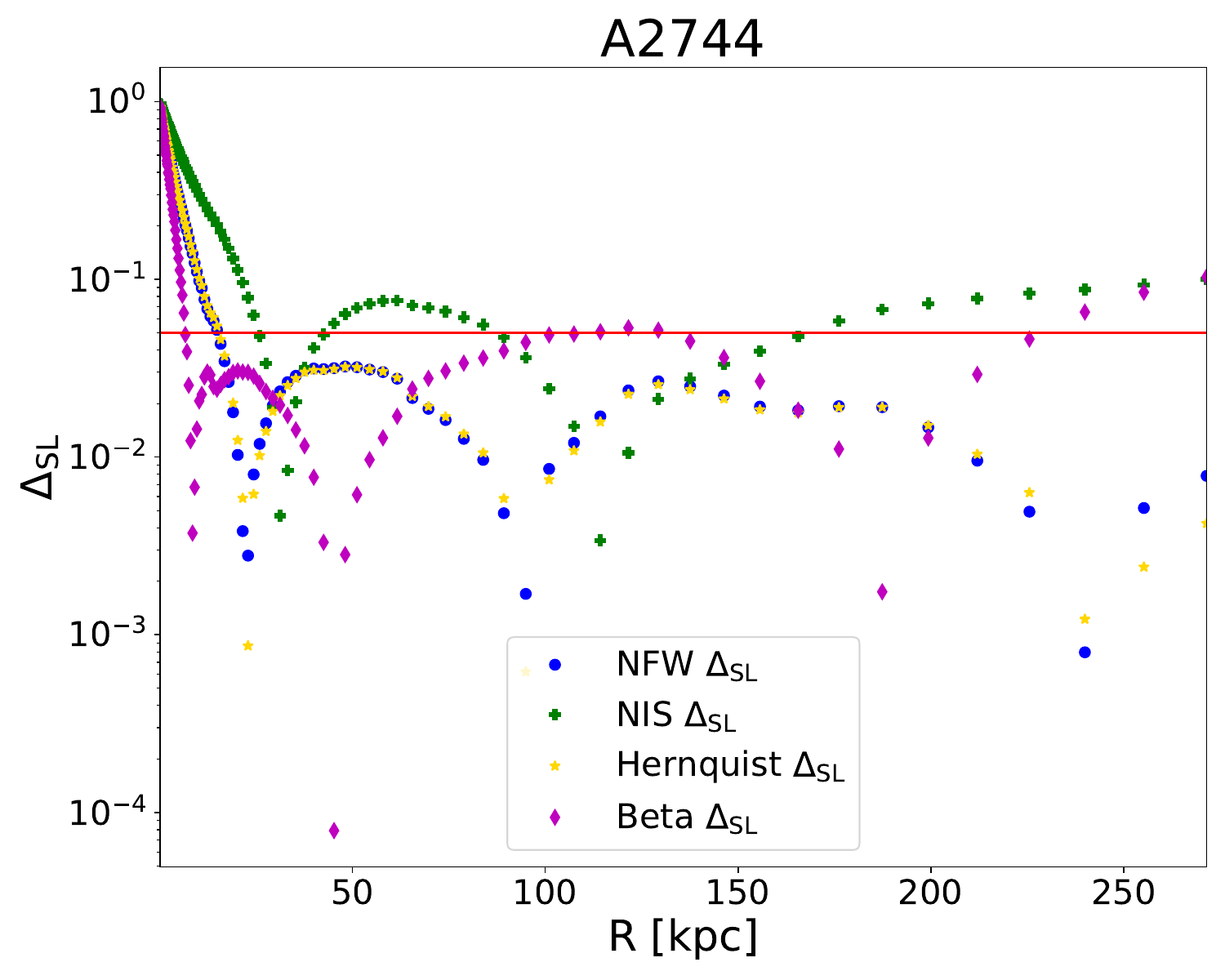}
\end{subfigure}
\begin{subfigure}
\centering 
\includegraphics[scale=0.24]{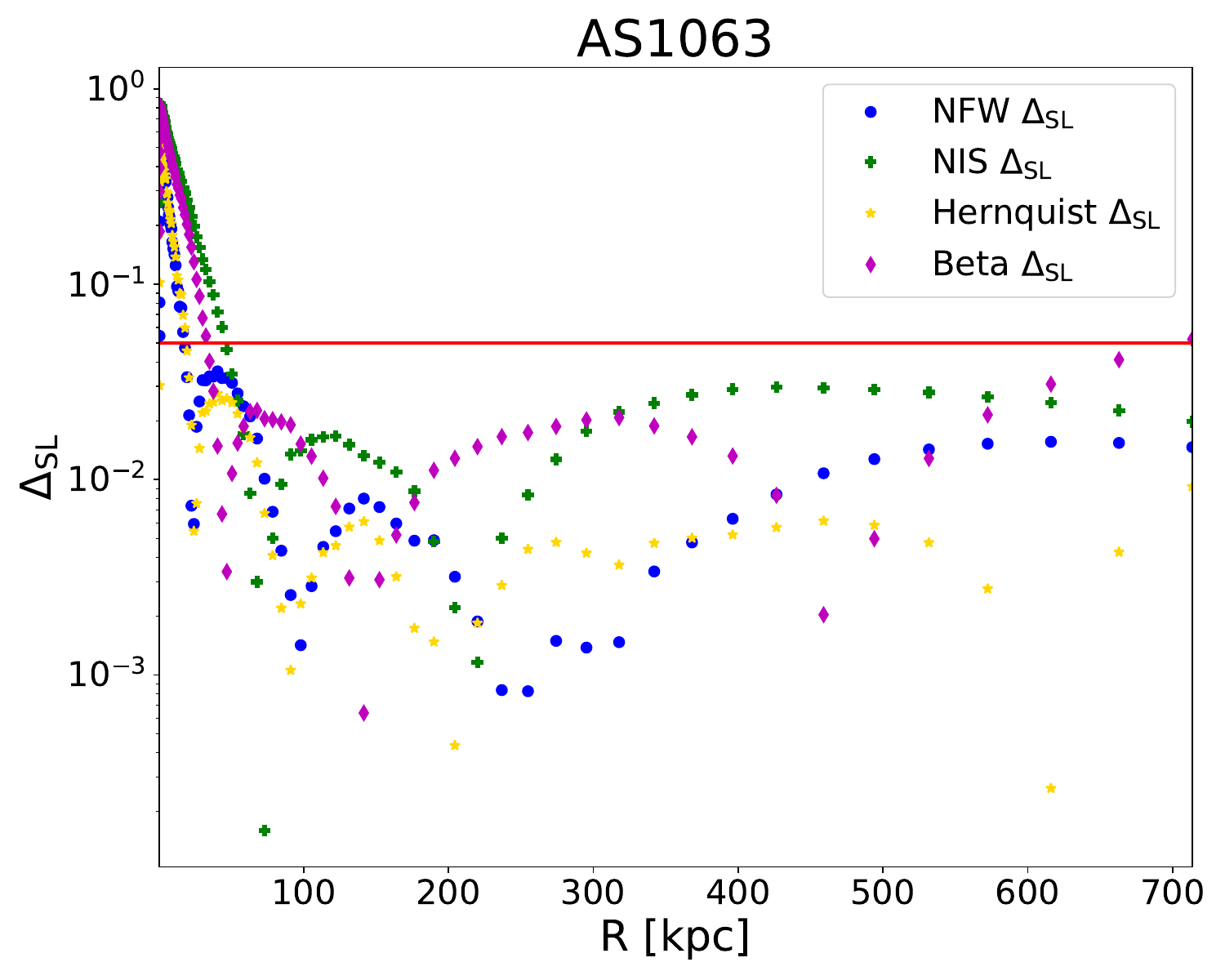}
\end{subfigure} \\
\begin{subfigure}
\centering 
\includegraphics[scale=0.24]{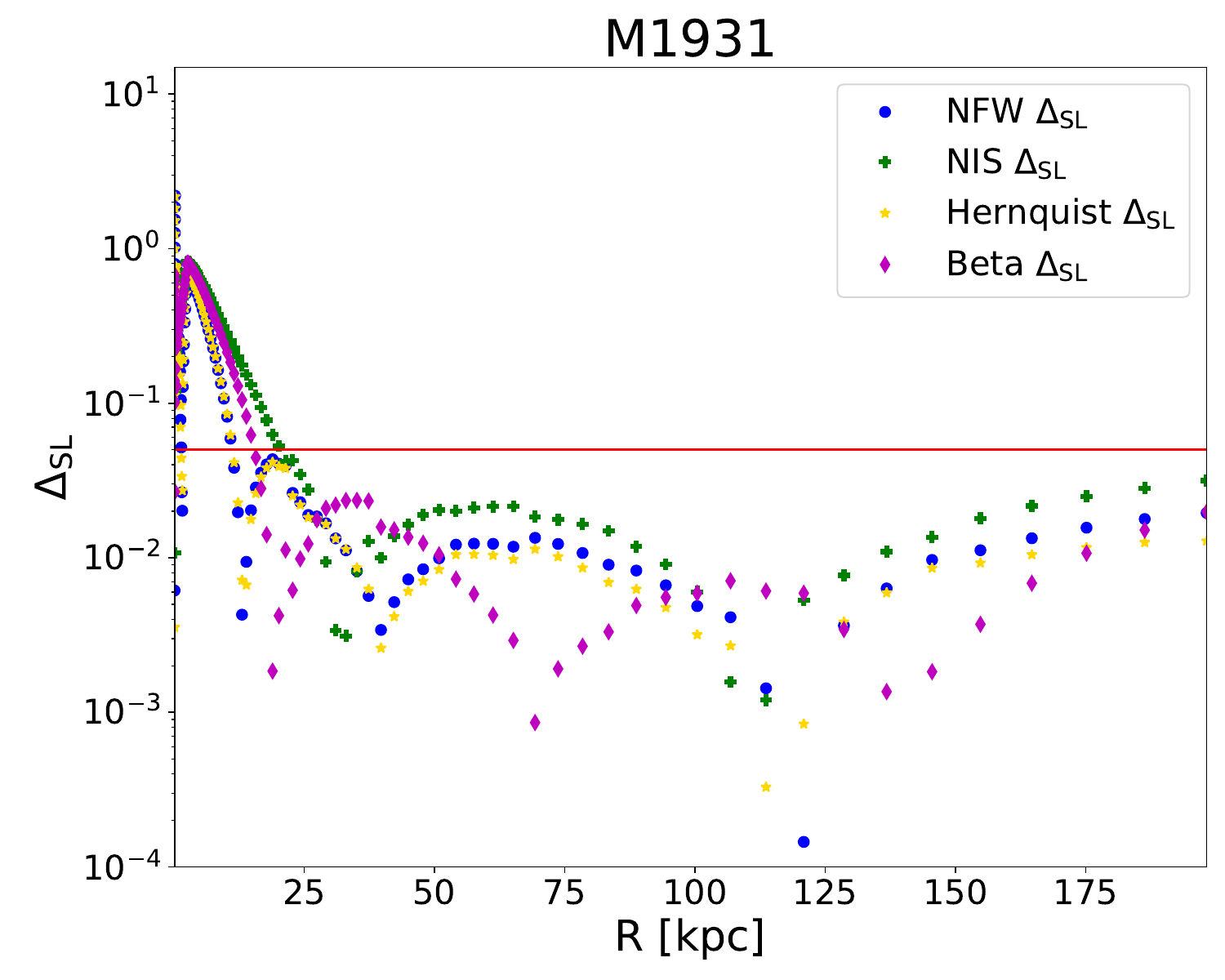}
\end{subfigure}
\begin{subfigure}
\centering 
\includegraphics[scale=0.24]{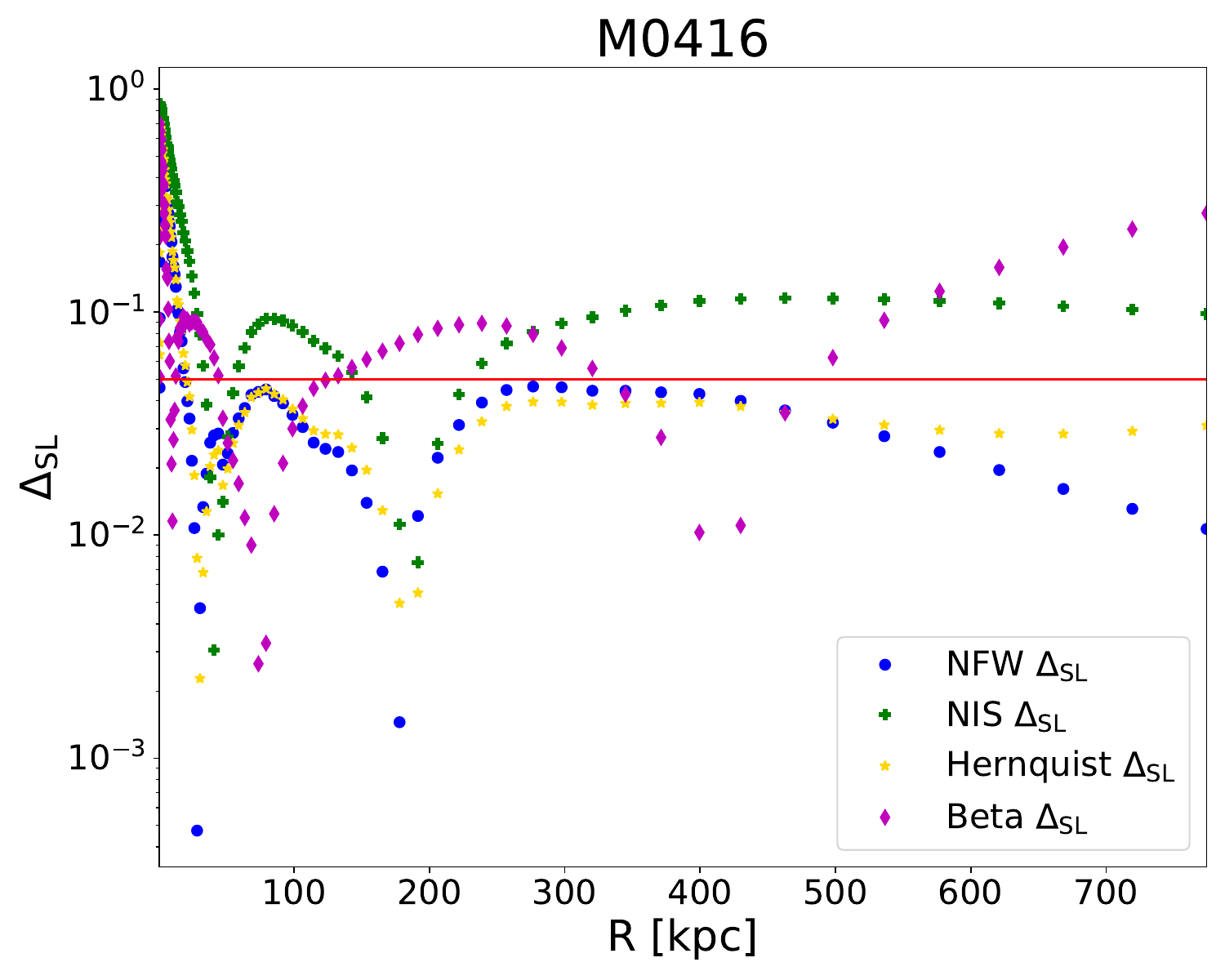}
\end{subfigure}
\begin{subfigure}
\centering 
\includegraphics[scale=0.24]{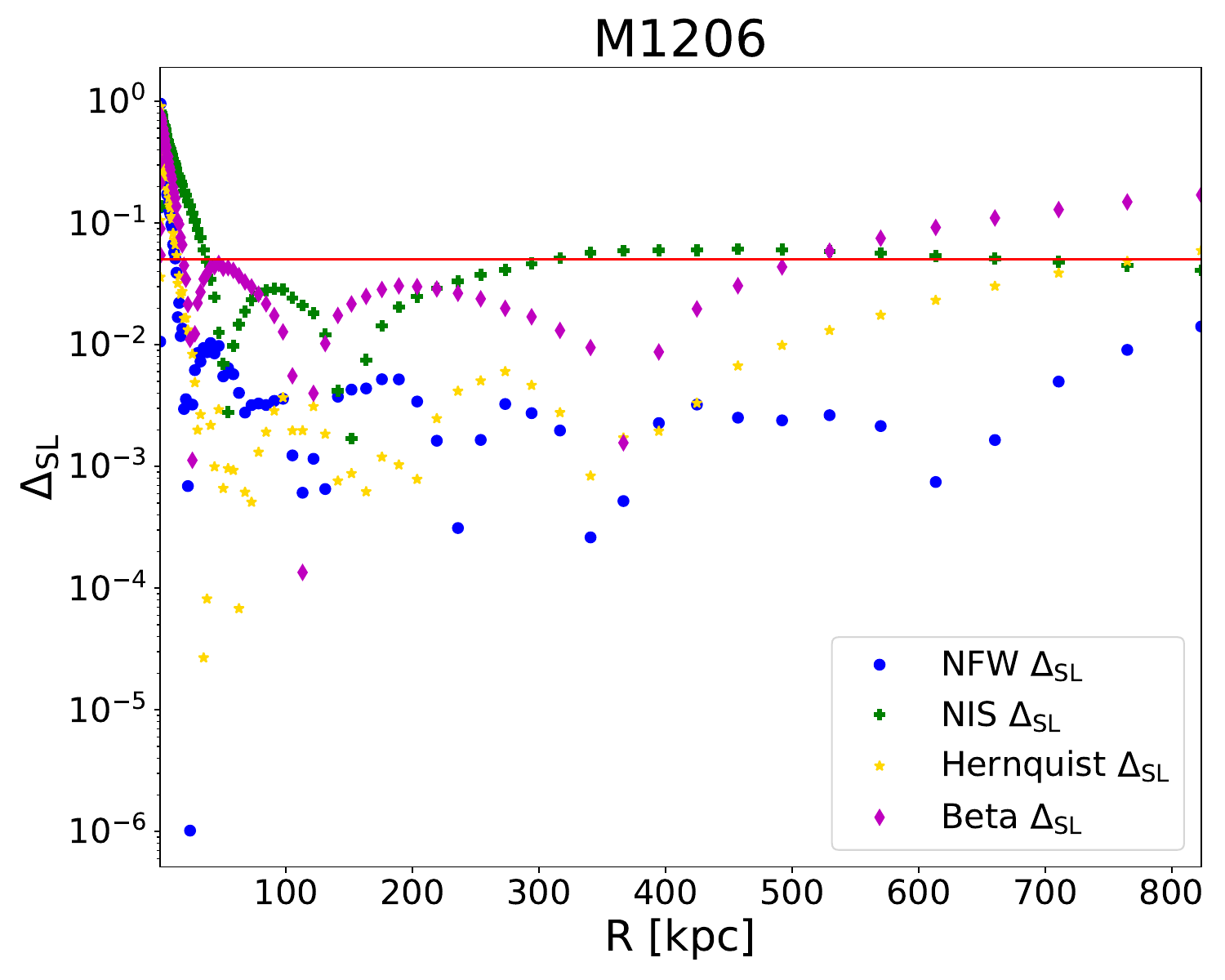}
\end{subfigure} \\
\begin{subfigure}
\centering 
\includegraphics[scale=0.24]{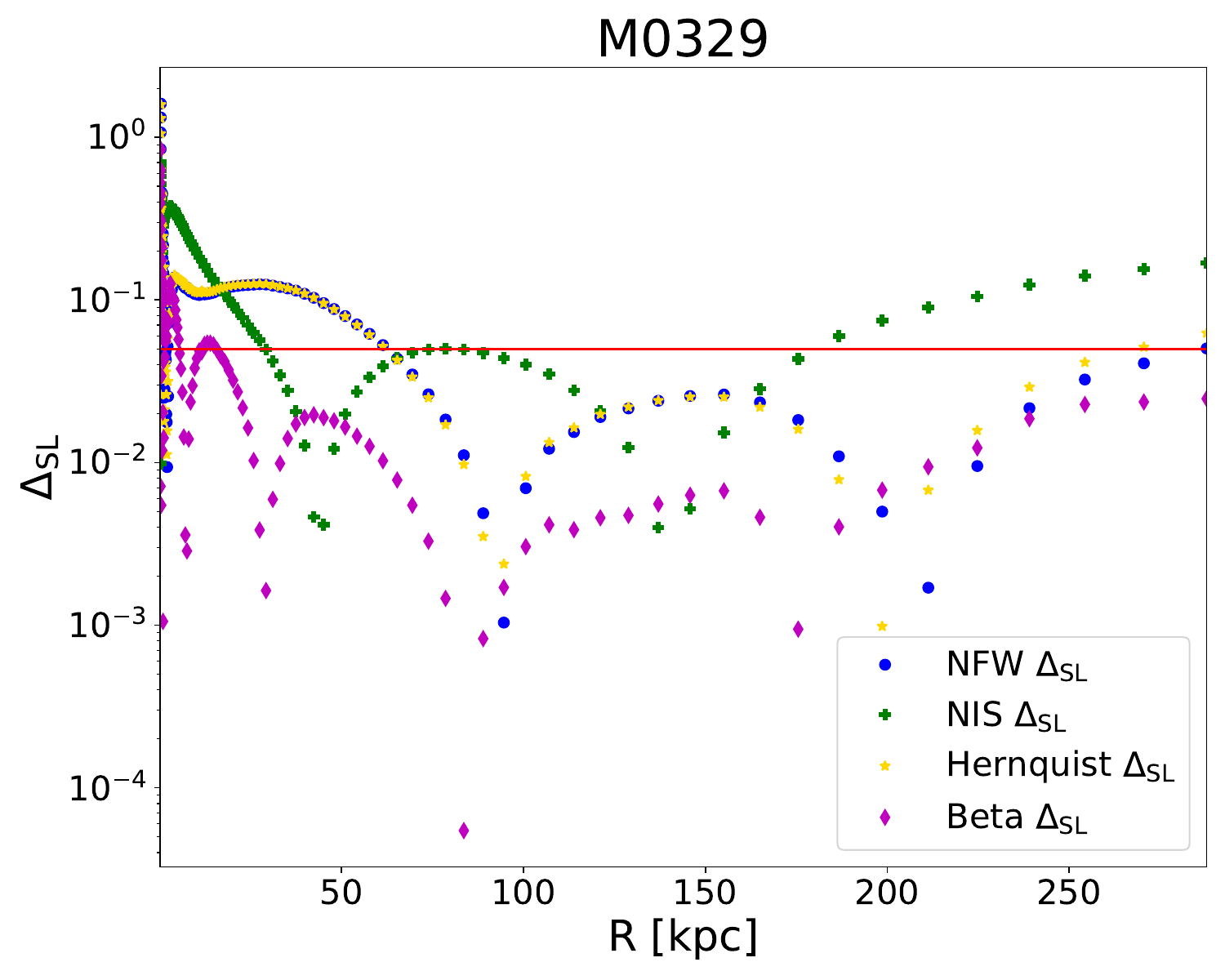}
\end{subfigure}
\begin{subfigure}
\centering 
\includegraphics[scale=0.24]{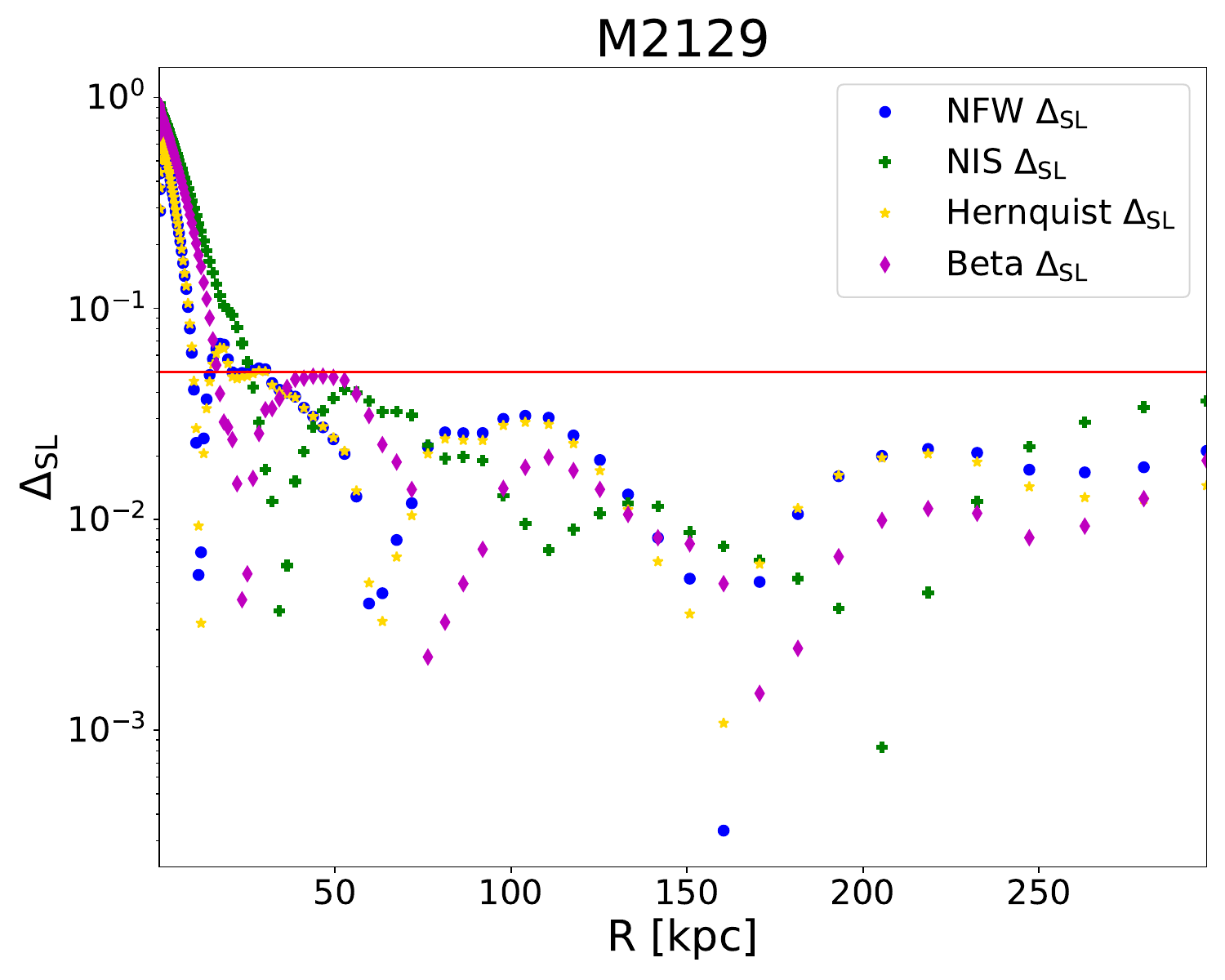}
\end{subfigure}
\begin{subfigure}
\centering 
\includegraphics[scale=0.24]{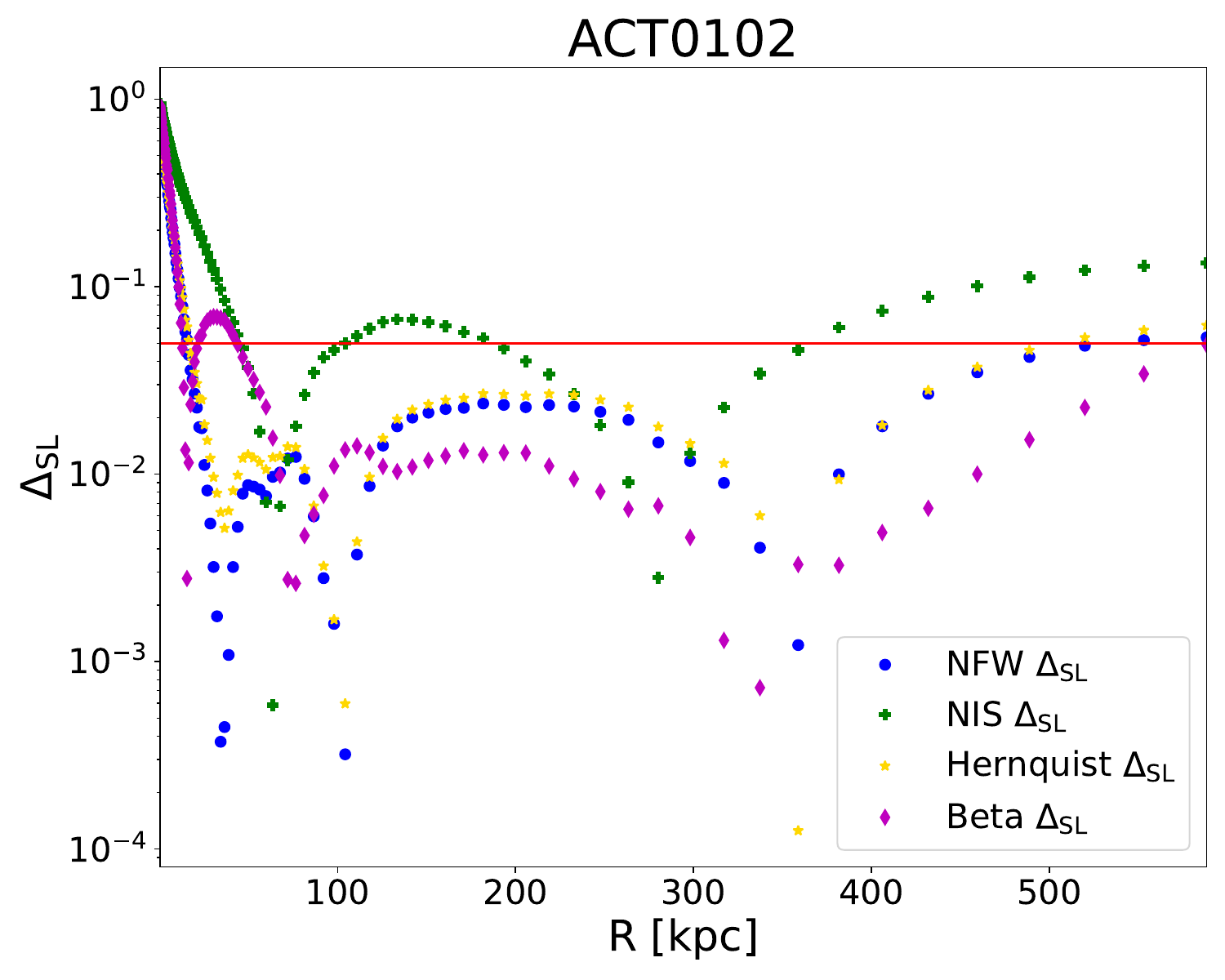}
\end{subfigure} \\
\caption{Absolute values of the relative differences between the measured and modeled cumulative projected total mass profiles rescaled to the measured profiles ($\Delta_\mathrm{SL}$, see Equation \ref{eq.relerr}) for each galaxy cluster. The red horizontal line in each plot represents the 5\% treshold.}
\label{fig.errrels}
\end{figure*}

We fit the measured $M_\mathrm{SL}(R)$ profiles with all the mass models we presented in Section \ref{sec.massmodels}: Figure \ref{fig.totprof} demonstrates that different mass profiles are equally good to reproduce the SL total mass measurements. We observe that focusing on the cluster central regions, the dPIE and NIS models produce very similar results. We can explain this behavior by noticing that the size of the investigated radial regions are small compared to the scale of a possible truncation radius $r_\mathrm{T}$ and that the dPIE profile reduces to a NIS profile if $r_\mathrm{T}\gg r$, like in our case. Henceforth, we report only the results for the NIS model, as they coincide with those of the dPIE model.
In Figure \ref{fig.errrels} we show the relative differences $\Delta_\mathrm{SL}$, defined as
\begin{equation}
\label{eq.relerr}
\Delta_\mathrm{SL}=  \left\vert\frac{M_\mathrm{SL}(R)-M_\mathrm{model}^\mathrm{2D}(R)}{M_\mathrm{SL}(R)}\right\vert,
\end{equation}
where $M_\mathrm{model}^\mathrm{2D}(R)$ is the 2D projection (see Equation \ref{eq.proj}) of the mass models listed in Section \ref{sec.massmodels}.
The $\Delta_\mathrm{SL}$ values are almost always small, except in the very central regions up to 20-25 kpc. Additionally, we report in Table \ref{tab.chi2} the values of the reduced $\chi^2$ for the different models. These values are a significantly larger than 1 for some of the galaxy clusters in the sample. This is due to the simplified (circular, one-component) total mass parametrization that we adopt. As explained in Section \ref{sec.stronglensdata}, the very small errors (barely visible in Figure \ref{fig.totprof}) affecting the measured projected total mass values result in very peaked probability distributions, and consequently small uncertainties, for the free parameter values of each model. Nevertheless, the small relative difference between the models and measurements (see Figure \ref{fig.errrels}) demonstrates that our models are good global probes of the measured total mass profiles studied herein. In the following subsections, we discuss the results of our fits and the performances of our mass models for each galaxy cluster in detail.

\begin{table}
\centering
\caption{Values of the reduced $\chi^2$ for each mass profile fitted to the sample of galaxy clusters}
\label{tab.chi2}
\begin{tabular}{ccccc} 
\toprule
Cluster & NFW & NIS & Hernquist & Beta \\
\midrule
RXJ2129 & 19.3 & 34.6 & 19.5 & 16.6\\
A2744 & 77.7 & 179.2 & 78.2 & 67.7\\
AS1063 & 18.6 & 81.9 & 17.1 & 58.3\\
M1931 & 1.8 & 3.0 & 1.7 & 1.9\\
M0416 & 40.4 & 124.3 & 41.9 & 46.9 \\
M1206 & 22.6 & 149.8 & 26.0 & 100.9\\
M0329 & 49.5 & 57.3 & 48.5 & 14.7\\
M2129 & 16.6 & 35.4 & 16.2 & 28.8\\
ACT0102 & 5.8 & 18.4 & 6.1 & 7.5\\
\bottomrule 
\end{tabular}
\end{table}

\begin{sidewaystable}
\caption{Median values and $1\sigma$ confidence level errors of the parameters for each total mass model for each galaxy cluster.}
\label{tab.paramtot}
\centering
\begin{tabular}{cccccccccc}
\toprule
\multirow{2}{*}{Cluster}  & \multicolumn{2}{c}{NFW} & \multicolumn{2}{c}{NIS} & \multicolumn{2}{c}{Hernquist} & \multicolumn{3}{c}{Beta} \\
\cmidrule(lr){2-3} \cmidrule(lr){4-5} \cmidrule(lr){6-7} \cmidrule(lr){8-10}
& $\rho_0$ & $r_\mathrm{S}$ & $\rho_0$ & $r_\mathrm{S}$ & $\rho_0$ & $r_\mathrm{S}$ & $\rho_0$ & $r_\mathrm{S}$ & $\beta$ \\
\midrule
RXJ2129 & $1.71^{+0.14}_{-0.15} \times 10^{6}$ & $3.11^{+0.22}_{-0.17} \times 10^{2}$ & $5.33^{+0.18}_{-0.17} \times 10^{8}$ & $2.64^{+0.06}_{-0.06} \times 10^{1}$ & $1.26^{+0.08}_{-0.08} \times 10^{7}$ & $5.25^{+0.24}_{-0.22} \times 10^{2}$ & $4.35^{+2.86}_{-1.89} \times 10^{9}$ & $3.01^{+1.73}_{-0.97} \times 10^{0}$ & $4.83^{+0.07}_{-0.05} \times 10^{-1}$ \\
A2744 & $6.35^{+0.24}_{-0.24} \times 10^{5}$ & $7.13^{+0.20}_{-0.19} \times 10^{2}$ & $2.29^{+0.04}_{-0.04} \times 10^{8}$ & $5.51^{+0.07}_{-0.07} \times 10^{1}$ & $4.63^{+0.17}_{-0.15} \times 10^{6}$ & $1.21^{+0.03}_{-0.03} \times 10^{3}$ & $9.34^{+14.26}_{-5.57} \times 10^{9}$ & $1.66^{+1.61}_{-0.81} \times 10^{0}$ & $4.66^{+0.02}_{-0.02} \times 10^{-1}$ \\
AS1063 & $9.38^{+0.05}_{-0.05} \times 10^{5}$ & $6.28^{+0.02}_{-0.02} \times 10^{2}$ & $1.80^{+0.02}_{-0.02} \times 10^{8}$ & $7.38^{+0.04}_{-0.04} \times 10^{1}$ & $6.37^{+0.03}_{-0.02} \times 10^{6}$ & $1.12^{+0.00}_{-0.00} \times 10^{3}$ & $2.52^{+0.06}_{-0.08} \times 10^{8}$ & $5.15^{+0.16}_{-0.11} \times 10^{1}$ & $6.13^{+0.03}_{-0.02} \times 10^{-1}$ \\
M1931 & $1.81^{+0.12}_{-0.15} \times 10^{6}$ & $3.72^{+0.22}_{-0.14} \times 10^{2}$ & $4.17^{+0.16}_{-0.20} \times 10^{8}$ & $3.90^{+0.12}_{-0.09} \times 10^{1}$ & $1.30^{+0.08}_{-0.08} \times 10^{7}$ & $6.41^{+0.27}_{-0.24} \times 10^{2}$ & $6.22^{+0.73}_{-0.62} \times 10^{8}$ & $2.36^{+0.28}_{-0.26} \times 10^{1}$ & $5.82^{+0.13}_{-0.13} \times 10^{-1}$ \\
M0416 & $6.12^{+0.08}_{-0.10} \times 10^{5}$ & $6.74^{+0.08}_{-0.06} \times 10^{2}$ & $1.61^{+0.04}_{-0.04} \times 10^{8}$ & $6.52^{+0.10}_{-0.10} \times 10^{1}$ & $4.17^{+0.05}_{-0.09} \times 10^{6}$ & $1.21^{+0.02}_{-0.01} \times 10^{3}$ & $2.59^{+1.89}_{-1.79} \times 10^{9}$ & $5.39^{+8.02}_{-1.78} \times 10^{0}$ & $5.07^{+0.18}_{-0.04} \times 10^{-1}$ \\
M1206 & $1.29^{+0.02}_{-0.02} \times 10^{6}$ & $5.08^{+0.05}_{-0.05} \times 10^{2}$ & $2.74^{+0.04}_{-0.03} \times 10^{8}$ & $5.58^{+0.04}_{-0.05} \times 10^{1}$ & $8.65^{+0.12}_{-0.13} \times 10^{6}$ & $9.15^{+0.08}_{-0.09} \times 10^{2}$ & $5.29^{+0.15}_{-0.15} \times 10^{8}$ & $2.89^{+0.07}_{-0.07} \times 10^{1}$ & $5.85^{+0.02}_{-0.02} \times 10^{-1}$ \\
M0329 & $1.34^{+0.05}_{-0.07} \times 10^{6}$ & $4.67^{+0.17}_{-0.12} \times 10^{2}$ & $5.23^{+0.19}_{-0.17} \times 10^{8}$ & $3.54^{+0.07}_{-0.08} \times 10^{1}$ & $9.68^{+0.39}_{-0.58} \times 10^{6}$ & $8.00^{+0.33}_{-0.20} \times 10^{2}$ & $1.81^{+0.10}_{-0.11} \times 10^{9}$ & $1.04^{+0.06}_{-0.05} \times 10^{1}$ & $5.43^{+0.03}_{-0.03} \times 10^{-1}$ \\
M2129 & $9.60^{+0.38}_{-0.40} \times 10^{5}$ & $5.71^{+0.16}_{-0.15} \times 10^{2}$ & $2.43^{+0.05}_{-0.05} \times 10^{8}$ & $5.66^{+0.07}_{-0.07} \times 10^{1}$ & $6.92^{+0.25}_{-0.24} \times 10^{6}$ & $9.82^{+0.23}_{-0.22} \times 10^{2}$ & $3.66^{+0.96}_{-0.26} \times 10^{8}$ & $3.33^{+0.28}_{-0.76} \times 10^{1}$ & $5.77^{+0.10}_{-0.26} \times 10^{-1}$ \\
ACT0102 & $2.49^{+0.19}_{-0.17} \times 10^{5}$ & $1.38^{+0.07}_{-0.06} \times 10^{3}$ & $6.53^{+0.35}_{-0.33} \times 10^{7}$ & $1.33^{+0.05}_{-0.04} \times 10^{2}$ & $1.74^{+0.14}_{-0.12} \times 10^{6}$ & $2.42^{+0.11}_{-0.12} \times 10^{3}$ & $3.20^{+0.78}_{-0.60} \times 10^{8}$ & $2.70^{+0.50}_{-0.45} \times 10^{1}$ & $5.18^{+0.06}_{-0.06} \times 10^{-1}$ \\
\bottomrule
\end{tabular}
\tablefoot{The total mass density scales are expressed in $\mathrm{M_\odot /kpc^3}$ and the length scales in kpc.}
\end{sidewaystable}

\subsection{M0416, M1206, and ACT0102} \label{sec.NFWclusters}

The galaxy clusters M0416, M1206, and ACT0102 are those that are best represented by a NFW mass profile. Interestingly, these three clusters share another common feature: they are all proven to be merging clusters (e.g. \citealt{Jee_2014_ElGordoWL, Balestra_2016_M0416}). As we show in Figure \ref{fig.totprof}, while for small radii all models reproduce well the $M_\mathrm{SL}(R)$ profiles, the NFW model fits better in the outer regions of the three galaxy clusters. This behavior is confirmed by the reduced $\chi^2$ values reported in Table \ref{tab.chi2}. We also note that the results obtained from the fits of the Hernquist profile are almost as compatible with the data as the NFW ones. This demonstrates that such profile is also equally good to reconstruct the clusters total mass profiles.

\subsection{AS1063, M1931, and M2129} \label{sec.HERclusters}

The model that best describes the total mass profile of AS1063, M1931, and M2129, as we show in Table \ref{tab.chi2}, is the Hernquist profile. For these three galaxy clusters, the difference between the proposed models is not clearly visible on the plots in Figure \ref{fig.totprof}. Although the Hernquist model is the best one for these clusters, the small difference between the Hernquist and NFW reduced $\chi^2$ values make these two models equally good in the fit to the measurements.

We note an interesting feature in the measured total mass profile of M1931: at very small radii, below 5 kpc, there is a kink. This peculiar characteristic is due to the underlying model of the DM halo that was obtained in the strong lensing analysis (see Section \ref{sec.stronglensdata}). For most of the galaxy clusters, the reconstructed DM mass distribution is approximately flat in the center, however the main DM halo of M1931 (similarly to AS1063 and M2129) has a particularly small core radius, unlike AS1063 and M2129. Hence, a small difference between the DM halo and the BCG centers is present, and in the case of M1931 is particularly visible.

\subsection{RXJ2129, A2744, and M0329} \label{sec.BETAclusters}

For RXJ2129, A2744, and M0329, the best-fit of their $M_\mathrm{SL}(R)$ is obtained with the beta model. The strong agreement between this model and the measured total mass profile for M0329 is clearly visible in Figure \ref{fig.totprof}. Even if it is not evident from the figures, the beta profile is the best-fitting model also for RXJ2129 and A2744, as shown in Table \ref{tab.chi2}. We note that for these three clusters the other available models are not as competitive with the best one, as they were in the previous cases. Moreover, similarly to M1931, we find a kink in the projected mass profile of RXJ2129, at radii below 5-6 kpc. 

\section{Virial mass relation} \label{sec.scalinglaws}

\begin{figure*}
\centering
\includegraphics[scale=0.365]{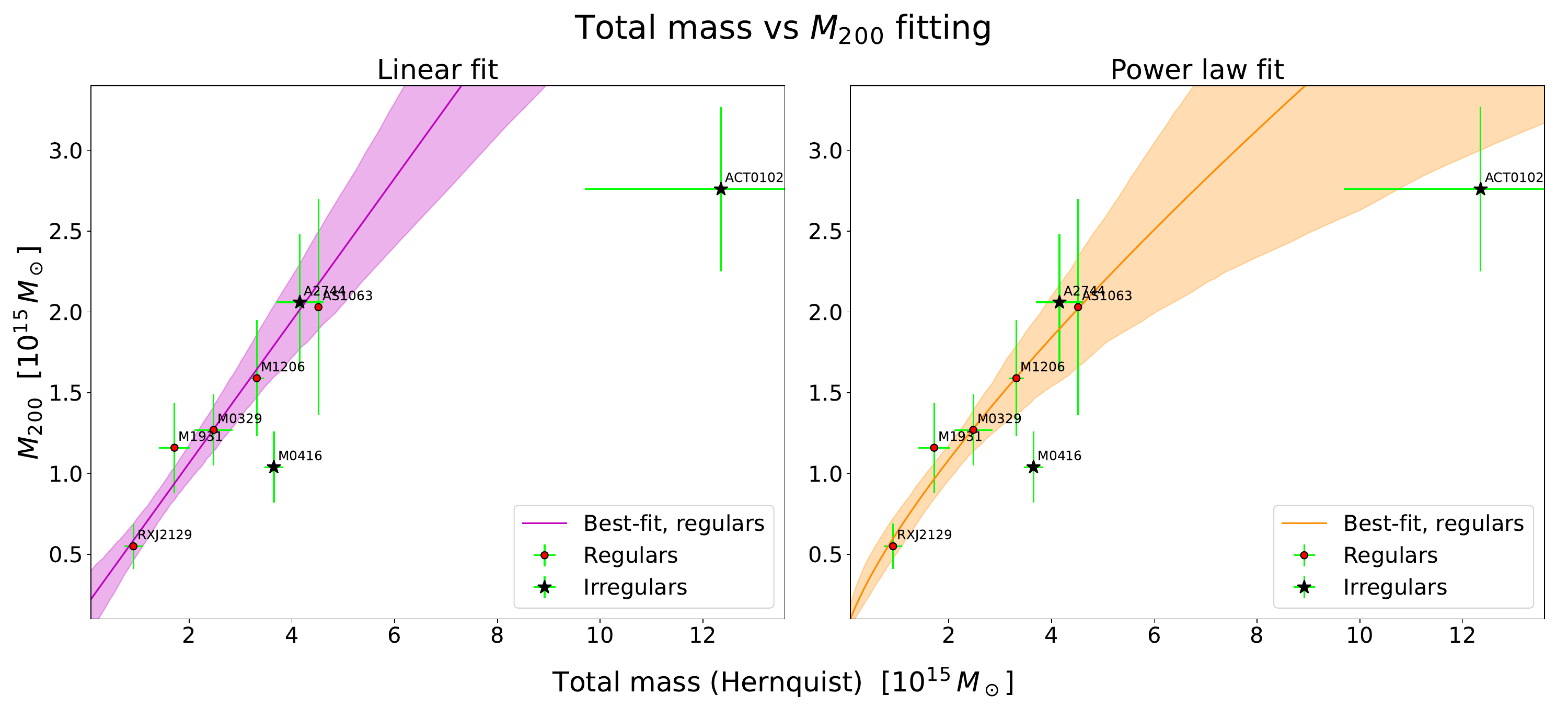}
\caption{Scaling relation between the total mass from the Hernquist profile, $ M_\mathrm{H}^\mathrm{tot}$, and the $M_{200\mathrm{c}}$ values from the literature \citep{Jee_2014_ElGordoWL, Medezinski_2016_A2744WL, Umetsu_2018_CLASHmassWL}. On the left, the solid line, with its corresponding $1\sigma$ shaded zone, is the linear best fit. On the right, instead, the solid line and its respective shaded zone is the power law best fit of the relation.}
\label{fig.m200rel}
\end{figure*}

Among the adopted mass profiles listed in Section \ref{sec.massmodels}, we note that the Hernquist profile is the only one that has a finite value of $M(r)$ for $r\rightarrow + \infty$. We call this quantity $M_\mathrm{H}^\mathrm{tot}$, which from Equation \ref{eq.her} is

\begin{equation}
    M_\mathrm{H}^\mathrm{tot} = M_\mathrm{H} (r\rightarrow + \infty)= \frac{\rho_\mathrm{0}r_\mathrm{s}^3}{2}.
\end{equation}
By using the inferred values and uncertainties of $\rho_\mathrm{0}$ and $r_\mathrm{s}$, we measure the values of $M_\mathrm{H}^\mathrm{tot}$ for each cluster in the sample and we compare them with the values of $M_{200\mathrm{c}}$ obtained from independent weak lensing (WL) analyses. Specifically, as shown in Table \ref{tab.clusters}, the available $M_{200\mathrm{c}}$ values are measured by \cite{Umetsu_2018_CLASHmassWL}\footnote{In the paper by \cite{Umetsu_2018_CLASHmassWL}, the authors employ a cosmological model slightly different from ours, setting $\Omega_\mathrm{m}=0.27$, $\Omega_\Lambda=0.73$, and $H_0 = 70 \ \mathrm{km \, s} ^{-1} \ \mathrm{Mpc}^{-1}$. However, we have checked that the effective change in the critical surface mass density $\Sigma_\mathrm{crit}$ in the WL analysis introduced by these different parameter values is $\lesssim 2\%$. Therefore, we neglect the errors related to this change, since they are significantly smaller than the statistical uncertainties reported by the authors.} for AS1063, M0416, M1206, RXJ2129, M1931, and M0329, \cite{Medezinski_2016_A2744WL} for A2744, and \cite{Jee_2014_ElGordoWL} for ACT0102. At the time of writing this paper, there are no available measurements for the $M_{200\mathrm{c}}$ value of M2129. Hence, we exclude this galaxy cluster from the following analyses.

\begin{table}
\centering
\caption{Values of the fitted parameters for each proposed scaling law.}
\label{tab.fitparamreg}
\begin{tabular}{cccc}
\toprule
Scaling law - Regulars & A & B & $\chi^2$ \\
\midrule
$AM_\mathrm{H}^\mathrm{tot}+B$ & $0.4^{+0.1}_{-0.1}$ & $0.2^{+0.2}_{-0.2}$ & 0.25 \\
$A(M_\mathrm{H}^\mathrm{tot})^B$ & $0.6^{+0.1}_{-0.1}$ & $0.8^{+0.2}_{-0.2}$ & 0.20 \\
$Ar_\mathrm{S}+B$ & $1.2^{+0.4}_{-0.4}$ & $1.1^{+0.3}_{-0.3} \times 10^3$ & 0.24 \\
\bottomrule 
\end{tabular}
\tablefoot{We present the values for the $M_\mathrm{H}^\mathrm{tot}$-$M_{200\mathrm{c}}$ scaling law and for the linear $r_\mathrm{S}$-$R_{200\mathrm{c}}$ relation in the "regulars only" galaxy cluster sample. In the rightmost column, we report the reduced $\chi^2$ values of each fit.}
\end{table}

In Figure \ref{fig.m200rel}, we plot $ M_\mathrm{H}^\mathrm{tot}$ as a function of $M_{200\mathrm{c}}$ for the different galaxy clusters. Moreover, we select a subsample of galaxy clusters, in which we include only the most regular ones (namely RXJ2129, AS1063, M1931, M1206, M0329, and M2129) from the original sample. Since our study is based on strong lensing models, we considered as "regular clusters" those that are modeled with a single main dark matter halo and do not present evident substructures. For these six clusters, indeed, the hypotesis of spherical symmetry is more valid, and we expect to obtain on this subsample a $ M_\mathrm{H}^\mathrm{tot}$ vs. $M_{200\mathrm{c}}$ scaling relation that is more precise than on the whole sample. We fit the $ M_\mathrm{H}^\mathrm{tot}$ and $M_{200\mathrm{c}}$ values of this smaller set of galaxy clusters with a linear function and a power law: in Table \ref{tab.fitparamreg}, we report the median values and $1\sigma$ confidence level errors of their free parameters, together with the corresponding reduced $\chi^2$ values. Despite some scatter, these scaling relations can provide a fair estimate of $M_{200\mathrm{c}}$ only by analyzing the innermost regions of a regular galaxy cluster.

\begin{figure}
\centering
\includegraphics[scale=0.36]{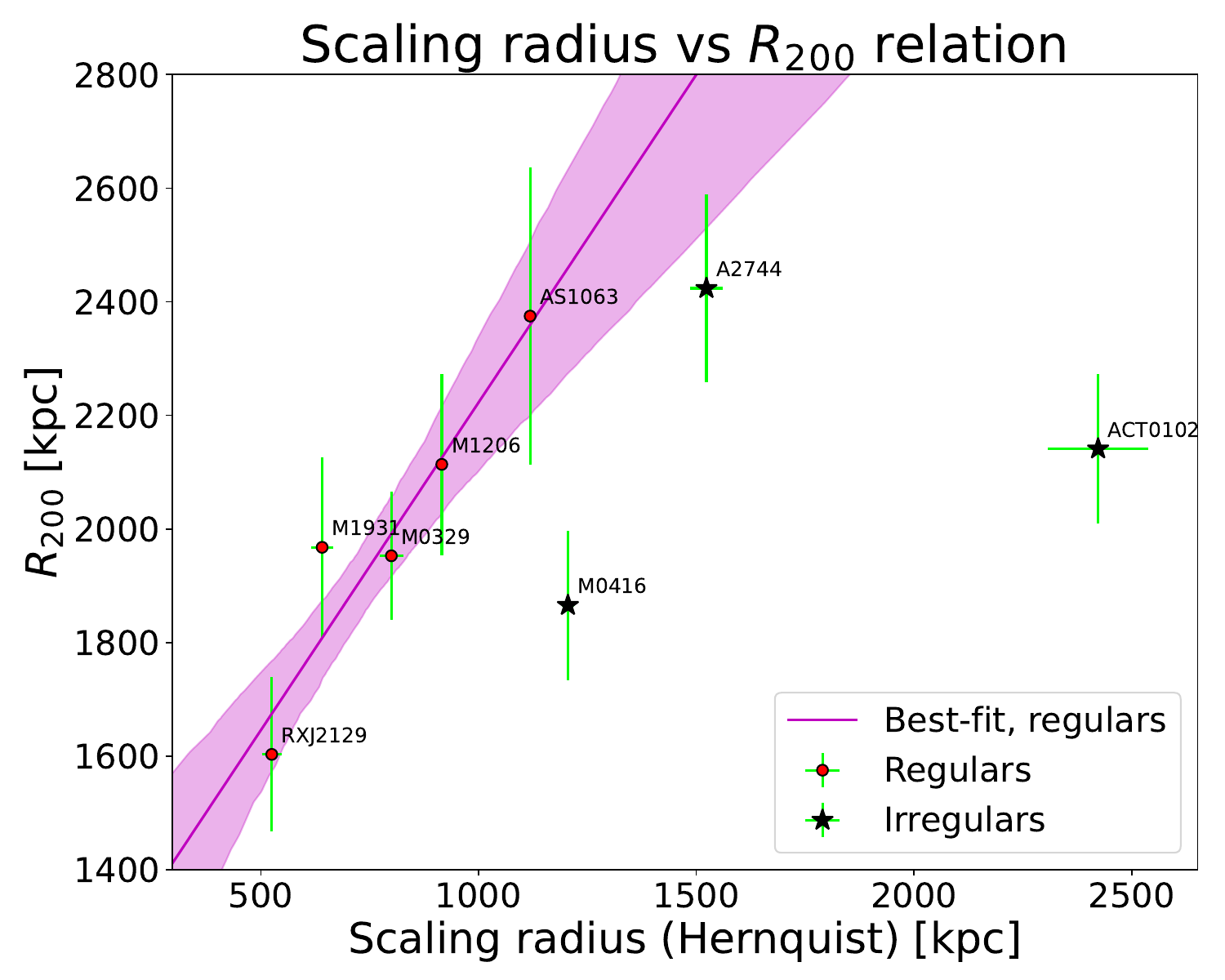}
\caption{Relation between the fitted scale radii of the Hernquist profile, $r_\mathrm{s}$, and the $R_{200\mathrm{c}}$ values obtained from the available values of $M_{200\mathrm{c}}$. The solid line is the linear best fit, with its corresponding $1\sigma$ uncertainty represented by the shaded zone.}
\label{fig.R200rel}
\end{figure}

Furthermore, we look for a possible linear scaling relation between the values of the characteristic radius of the Hernquist model ($r_\mathrm{S}$, see Table \ref{tab.paramtot}) and those of $R_{200\mathrm{c}}$. We computed the $R_{200\mathrm{c}}$ values and their errors starting from the $M_{200\mathrm{c}}$ measurements reported in Table \ref{tab.clusters}, by assuming the cosmological model described at the end of Section \ref{sec.intro}. In Figure \ref{fig.R200rel}, we plot the relation that we find between $r_\mathrm{S}$ and $R_{200\mathrm{c}}$. We note that here the difference between the two galaxy cluster subsamples is even clearer than in the previous case. We also observe that the $R_{200\mathrm{c}}$ value for ACT0102 is not the highest of the cluster sample, as could be expected since it has the highest $M_{200\mathrm{c}}$ value. This is due to the significantly higher cosmic critical mass density value at the ACT0102 redshift (0.87), that is $\sim 335 \, \mathrm{M_\odot /kpc^3}$, instead of the $\lesssim 200 \, \mathrm{M_\odot /kpc^3}$ for the other galaxy clusters.

We list the parameter values of this fit, and the reduced $\chi^2$ value in Table \ref{tab.fitparamreg}. In particular, we note that if we blindly apply the scaling law found from the regulars to the large scale radii of A2744, M0416, and ACT0102 (the less regular clusters), we find values of $R_{200\mathrm{c}}$ higher than those measured from WL. This behavior can be explained by noticing that the prominent substructures in A2744, M0416, and ACT0102 give an elongated shape to the projected total mass of these clusters (see \citealt{Bergamini_2019_RXCJ2248_M0416_M1206, Caminha_2023_ElGordoSL}), and in single-component models like ours these elongations translate into larger scale radii than in the regular case. The scatter on the $r_\mathrm{S}$-$R_{200\mathrm{c}}$ relation is similar to that of the $ M_\mathrm{H}^\mathrm{tot}$-$M_{200\mathrm{c}}$ one. However, also in this case, the observed scaling relation can provide useful information on virial quantities by extrapolating those obtained in the central regions.

\begin{figure}
\begin{subfigure}
\centering 
\includegraphics[scale=0.35]{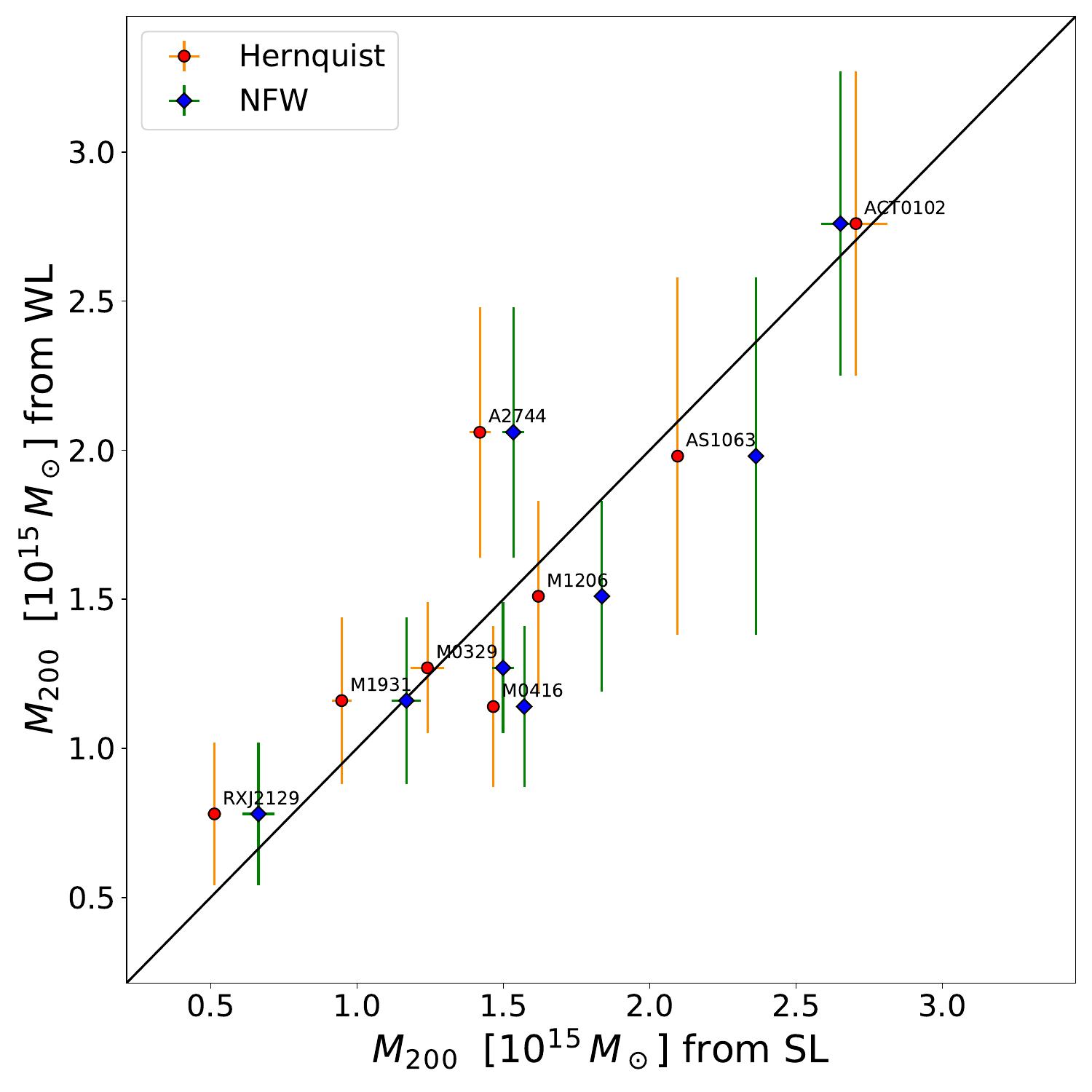}
\end{subfigure} \\
\begin{subfigure}
\centering 
\includegraphics[scale=0.35]{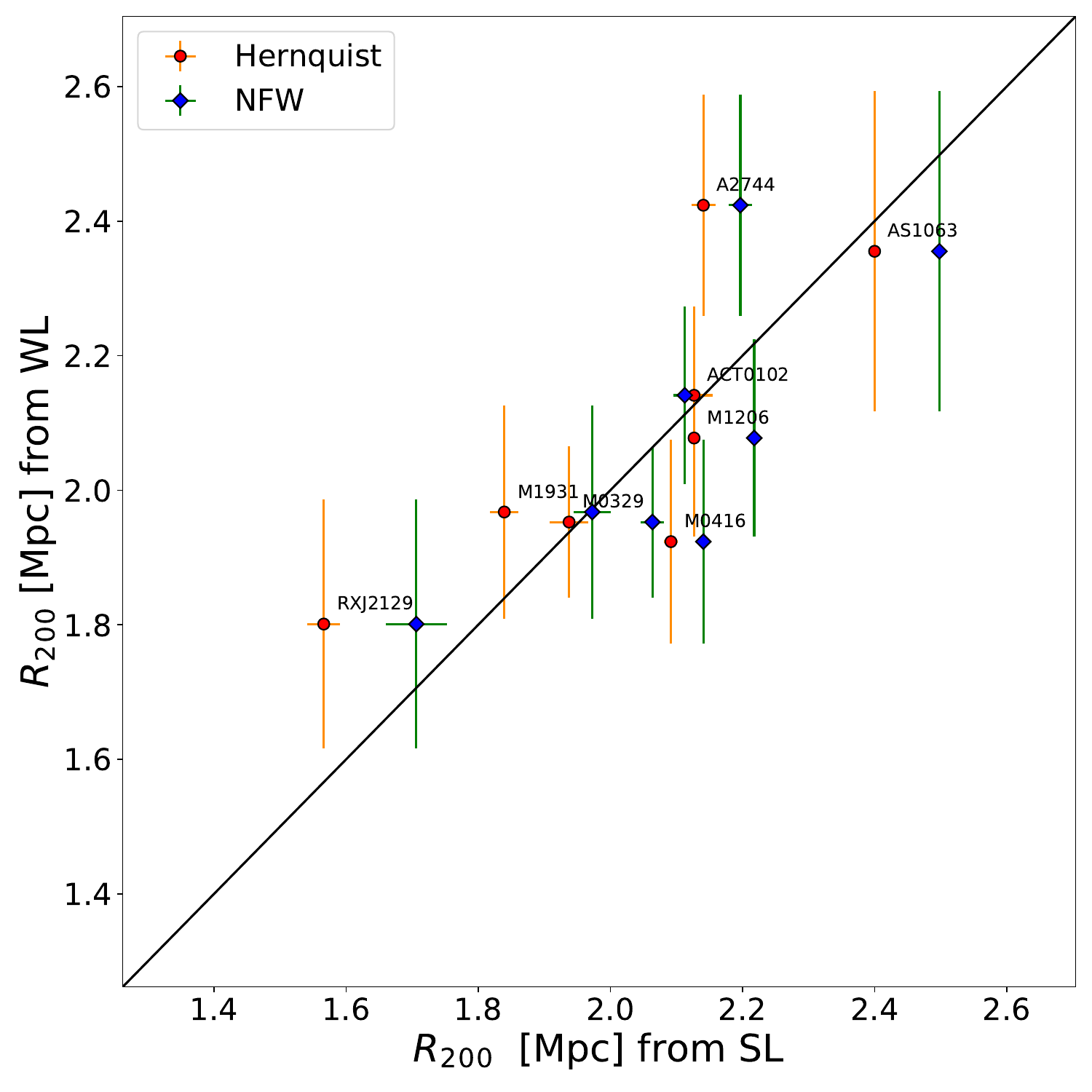}
\end{subfigure}
\caption{Upper panel: Comparison between the SL-based $M_{200\mathrm{c}}$ measurements and the WL-based ones. Lower panel: Comparison between the SL-based $R_{200\mathrm{c}}$ measurements and the WL-based ones. In both panels we note that these measurements are very close to have a 1:1 ratio, especially those relative to the most regular galaxy clusters.}
\label{fig.slwl200s}
\end{figure}

The relation between SL-based and WL-based measurements is further strenghtened by the fact that the $M_{200\mathrm{c}}$ and $R_{200\mathrm{c}}$ values measured in the weak-lensing regime are consistent with those that we can measure in our SL-based analyses. Indeed, if we consider the equation

\begin{equation}
\label{eq.rhocrit}
\frac{M_\mathrm{H}(R_{200\mathrm{c}})}{\frac{4}{3}\pi R_{200\mathrm{c}}^3}= 200 \rho_\mathrm{crit}(z),
\end{equation}
where $\rho_\mathrm{crit}(z)$ is the critical density of the Universe at the redshift $z$, and the fact that we fit the scale parameters $\rho_\mathrm{0}$ and $r_\mathrm{s}$, we can solve Equation \ref{eq.rhocrit} for $R_{200\mathrm{c}}$. Consequently, we can also compute 

\begin{equation}
\label{eq.m200sl}
M_{200\mathrm{c}}=\frac{4}{3}\pi (R_{200\mathrm{c}})^3 \times 200 \rho_\mathrm{crit}(z).
\end{equation}
Thus, we can measure the values of $R_{200\mathrm{c}}$ and $M_{200\mathrm{c}}$ starting from our SL-based mass fits. In Figure \ref{fig.slwl200s}, we compare these values derived from the fitted NFW and Hernquist profile with the WL results.

We note that the $R_{200\mathrm{c}}$ and $M_{200\mathrm{c}}$ values that we obtain from the results of our fits, especially those that are derived with the Hernquist profile, are very close to the weak lensing measurements in the literature \citep{Jee_2014_ElGordoWL, Medezinski_2016_A2744WL, Umetsu_2018_CLASHmassWL}. Conversely, the $M_{200\mathrm{c}}$ and $R_{200\mathrm{c}}$ values obtained from the NFW profiles are slightly overestimated with respect to the Hernquist ones. We also observe that the less regular clusters, i.e. A2744, M0416, and ACT0102, have a very good correspondence between the SL and WL measurements, despite the different values of their scale radius (see Section \ref{sec.scalinglaws}). This is because the larger scale radii are compensated by smaller central densities, following the behavior that we show in Figure \ref{fig.corrparam}, and discuss in Section \ref{sec.projeffects}. Hence, this balance between the parameters makes the $R_{200\mathrm{c}}$ and $M_{200\mathrm{c}}$ values of the less regular galaxy clusters as valid as those of the more regular ones. Overall, the excellent matching between the SL and WL measured $M_{200\mathrm{c}}$ values shows that high-quality SL models, with low RMS and tens of multiple images, broaden the information extraction capabilities to regions well beyond the borders of the multiple image positions.

\section{Discussion} \label{sec.discussion}

The datasets at our disposal, as explained in Section \ref{sec.stronglensdata}, consist in cumulative projected total mass profiles. The lack of information about the third spacial dimension has a twofold inconvenience: the first aspect concerns the real geometry of the structure that we study, making the assumptions for the models not completely verifiable; second, the stacking of information due to the projection operation may soften some features in the projected mass profile, such as mass substructures.

\subsection{The spherical symmetry assumption}

In this section we discuss how the assumption of spherical symmetry for the clusters influences the results we obtain. The hypothesis of spherical symmetry has the great advantage of resulting in simple and analytical solutions to all the deprojected mass models listed in Section \ref{sec.massmodels}. However, it excludes a possible (and often observed, e.g. \citealt{Despali_2014_tiaxialityclusters,Bonamigo_2015_triaxialityclusters}) bi/triaxial configuration. Moreover, the spherical assumption does not take into account the specific characteristics of galaxy clusters, such as the presence of mass substructures, which may introduce a "perturbation" in the total mass distribution and that is neglected in our hypothesis. Overall, we consider the spherical symmetry approximation a good tradeoff between its limitations and advantages, because its simplicity allows us not only to have analytical solutions, but also to handle a limited number of free parameters, which can be immediately compared with the measurements or with other models (as we do for the $M_\mathrm{H}^\mathrm{tot}$-$M_{200\mathrm{c}}$ and $r_\mathrm{S}$-$R_{200\mathrm{c}}$ relations). 

The impact of the spherical symmetry assumption is hard to quantify and it is different from cluster to cluster. However, we can quantify the impact of this hypothesis with two different tests. First, we can look at the reduced $\chi^2$ values presented in Table \ref{tab.chi2}. As we mentioned in Section \ref{sec.risultati}, these values are often significantly higher than 1. This could suggest that spherical models are too simplicistic in order to accurately represent the projected total mass profiles. The very small error bars on the measured quantities are responsible for the high values of the $\chi^2$. We also note that we are modeling, under the spherical symmetry assumption, some galaxy clusters that are known to be composed by more complex mass substructures (see Section \ref{sec.massmodels}).
However, the performances of the spherical models are particularly good in terms of relative errors. In Figure \ref{fig.errrels}, we illustrate the absolute values of the relative errors between the projected total mass measured from the SL analysis and our models. It is clear that for more than 90\% of the considered radii, the relative errors $\delta_\mathrm{R}$ (at least of the best-fitting model) are below the 5\% threshold (drawn as the red horizontal lines). This fact shows that, although the reduced $\chi^2$ values might suggest that the spherical models are not very accurate, the relative difference in measured and modeled total mass is particularly small. For these reasons, we consider our models as a competitive solution for the interpretation of the projected total mass profiles in the selected sample of galaxy clusters.

\subsection{Projection effects} \label{sec.projeffects}

In this section, we discuss how the projection of the total mass distribution, resulting from SL analysis, influences the characteristics of the final profiles. As shown by \cite{Bonamigo_2017_M0416,Bonamigo_2018_RXCJ2248_M0416_M1206}, some clusters studied in this work present evident mass substructures in their cores. In those works, the presence of massive substructures is taken into account in the strong lensing analyses. Except for a kink in the innermost regions (see Figure \ref{fig.totprof} and Section \ref{sec.NFWclusters}, \ref{sec.HERclusters}, \ref{sec.BETAclusters}), the cumulative total mass profiles do not show evident features or irregularities due to the presence of substructures. This behavior has to be reconducted to the nature of the SL analysis, which measures the total mass enclosed within a certain area (and so within a projected radius), summing all the relevant mass contributions on the projected plane. Consequently, the mass perturbations introduced by the substructures become less relevant and they do not show up in the cumulative projected total mass profiles. This argument made us safely neglect the presence and the effects of any substructure in our sample of galaxy clusters.

\begin{figure}
\centering
\includegraphics[scale=0.49]{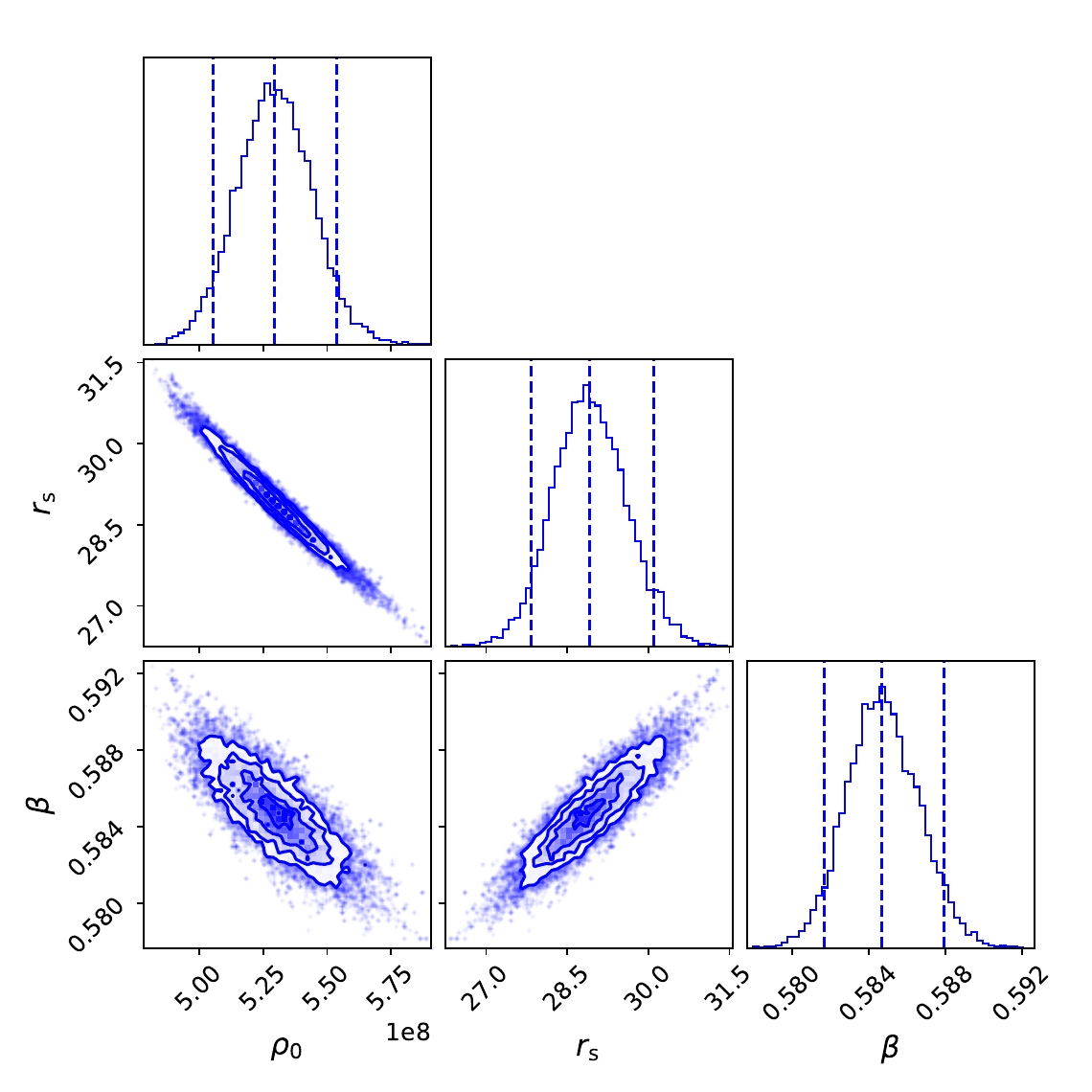}
\caption{Corner plot of the beta model parameters for the galaxy cluster M1206. The vertical, dashed lines in the histograms represent, from the left to the right, the 16th, 50th, and 84th percentiles of the distribution, while in the 2D histograms the 0.5, 1, 1.5, and 2-$\sigma$ equivalent contours are drawn. The central density $\rho_0$ and the scaling radius $r_\mathrm{s}$ show a strong correlation. This behavior is observed similarly for all models of each galaxy cluster.}
\label{fig.corrparam}
\end{figure}

Dealing with projected quantities has also an impact on the values of parameters of the adopted models. We notice indeed that two of the free parameters of the chosen mass models present a strong correlation (in the case of models with three free parameters there are not preferential couples of correlated parameters). As an example, we show in Figure \ref{fig.corrparam} the parameter degeneracies that we obtain when we fit the total mass profile of M1206 with the beta model. The values of the central density $\rho_0$ and the scaling radius $r_\mathrm{s}$ are strongly correlated. A similar behavior is found in every plot of the posterior distributions. The degeneracy between $\rho_0$ and $r_s$ for these models, indeed, is harder to break when some information is lacking due to the projection.

\subsection{Robustness tests} \label{sec.robtests}

As mentioned in Section \ref{sec.stronglensdata}, we test if the values of best-fit parameters of the mass models change significantly when the center of the circular regions is moved to a different point, and the total mass distribution within the new regions is consequently fitted. We also test if the values of the best-fit parameters change appreciably by increasing the maximum radii of the circular areas, while mantaining fixed the centers. 

\begin{figure}
\centering
\includegraphics[scale=0.36]{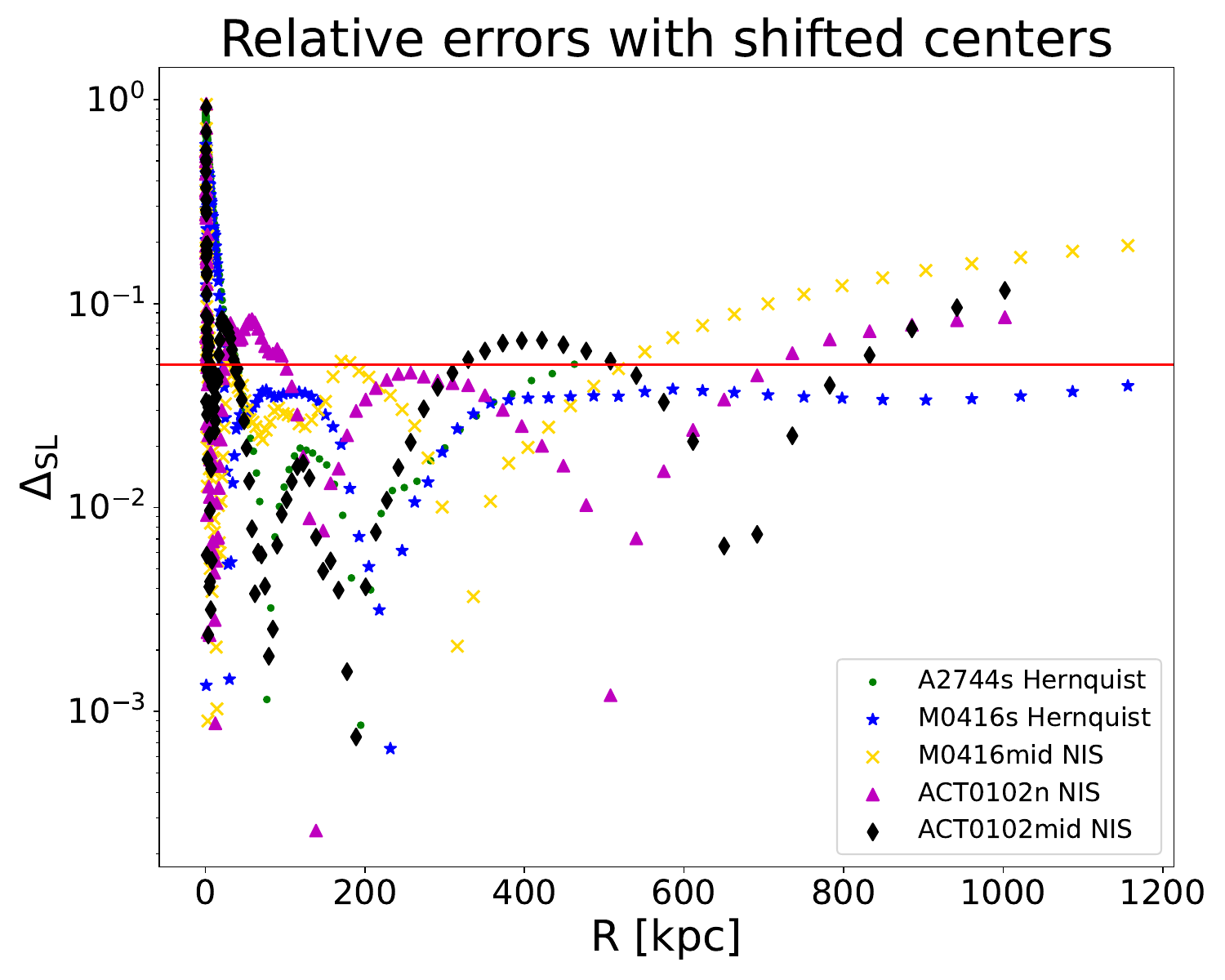}
\caption{Relative differences $\Delta_\mathrm{SL}$ (absolute values) of the mass models fitted on the projected profiles with shifted centers. As in Figure \ref{fig.errrels}, the red horizontal line in the plot is the 5\% treshold. Although the parameters of the models may change if the coordinates of the center point change, we note that such a change does not affect significantly the accuracy of spherical models in reproducing the corresponding projected mass profiles, with relative differences mostly below the 5\%.}
\label{fig.relmoving}
\end{figure}

We choose to change the centers of A2744, M0416 and ACT0102, since they are the least regular clusters in our sample. Specifically, for A2744, we measure a new total mass profile centered on the southern BCG of the galaxy cluster (cfr. BCG-S in \citealt{Bergamini_2023_A2744SL}), contrary to the original one that was centered on the BCG-N; for M0416, we measure two new total mass profiles, one centered on the southern BCG, while the other one located in the mean point between the two BCGs; for ACT0102, we measure two new profiles centered in the middle of the northern clump\footnote{Since in the northern region of ACT0102 there is no evident BCG, we assigned the center of this northern clump, namely at RA 1:02:53.2, DEC $-$49:15:09.7, between two bright cluster members.} and on the mean point between this latter position and the BCG. We use the Hernquist model, when the profiles are centered on a BCG, and the NIS model for the other positions. The choice of testing only the Hernquist model is justified by the large computational times required to explore all the models, and because this is the model adopted to calibrate the scaling laws in Section \ref{sec.scalinglaws}. On the other hand, we decide to test the NIS model in the BCG mid-points of M0416 and ACT0102 because at those locations the most relevant mass component is the large-scale DM halo of the galaxy clusters, characterized by a nearly flat profile in its core. We report in Table \ref{tab.movcentertest} the results for the original projected mass fits and the outcomes of the tests on the new regions for A2744, M0416, and ACT0102. Since we adopt different models for different center positions, we include in Table \ref{tab.movcentertest} the total mass value within 500 kpc computed with the best-fit parameter values. The distance of 500 kpc is the optimal compromise between the innermost regions of the clusters (where the multiple images are detected) and the outer regions, where the total mass values that we can obtain from our fitted models are fully based on an extrapolation. By comparing the results we observe a good agreement between the different models, thus indicating that the choice of the center position is not among the most relevant factors to determine the shape of the cumulative total mass profiles. Moreover, as we show in Figure \ref{fig.relmoving}, the relative differences $\Delta_\mathrm{SL}$ of the new fits with the shifted centers are still safely below the 5\% treshold for the vast majority of the considered radii. We can conclude that when the centers of the projected total mass profiles are moved, even if there can be some changes in the parameter values, our spherical models reproduce cumulative total mass values consistent with previous results.

We choose to do the "Increasing radius" test for M0329 and A2744, as the first one is representative of the regular galaxy cluster class (in which we count RXJ2129, AS1063, M1931, M1206, M0329, and M2129) and the second of the less regular class (in which we list A2744, M0416 and ACT0102). We perform this test in two ways: for A2744, we test all our 4 best models (NFW, NIS, Hernquist and Beta) assuming a maximum radius of $190\arcsec$ from the BCG-N, so that the resulting area encloses also the more external clumps G1, G2, and G3 (see \citealt{Bergamini_2023_A2744SL} for reference); for M0329, instead, we change 5 times the maximum radius, while keeping fixed the Hernquist model.

As we show in Table \ref{tab.radincrtest}, for M0329 the growth of the maximum radius does not affect particularly the parameters values, even if the considered area is 36 times larger than the original one. This behavior is reflected into the total mass values within 500 kpc, that do not change significantly when the maximum radius is increased. A2744 presents a more complex situation, in which the parameter values show a variation with respect to the original choice of the mass profile center. This is due to the fact that a non-regular cluster, like A2744, at large radii can have local peaks in the mass distribution (in the case of A2744, these peaks correspond to the clumps G1, G2, and G3), that can significantly influence the behavior of the total mass fits. Consequently, the total mass values within 500 kpc can result inconsistent, although the relative difference between them lays below 15\%. However, we note that the variation of the parameters follows the correlation pattern described in Section \ref{sec.projeffects}: an increase of the central density counterbalanced by a decrease of the scaling radius. This behavior, but in a less evident way, can be also observed for M0329. 

\begin{table*}
\centering
\caption{Results of the moving center tests.}
\label{tab.movcentertest}
\begin{tabular}{cccccc} 
\toprule
\midrule
A2744 & Mass model & $\rho_0$ & $r_\mathrm{S}$ & $M(<500 \, \mathrm{kpc})$ & $\chi^2$ \\
\midrule
BCG-N & Hernquist & $4.63^{+0.17}_{-0.15} \times 10^{6}$ & $1.21^{+0.03}_{-0.03} \times 10^{3}$ & $3.5^{+0.2}_{-0.2} \times 10^{14}$ & 78.2 \\
BCG-S & Hernquist &$6.89^{+0.21}_{-0.22} \times 10^{6}$ & $9.09^{+0.20}_{-0.18} \times 10^{2}$ & $3.3^{+0.2}_{-0.2} \times 10^{14}$ & 48.9\\
\midrule
\midrule
M0416 & Mass model & $\rho_0$ & $r_\mathrm{S}$ & $M(<500 \, \mathrm{kpc})$ & $\chi^2$ \\
\midrule
BCG-N & Hernquist &$4.17^{+0.05}_{-0.09} \times 10^{6}$ & $1.21^{+0.02}_{-0.01} \times 10^{3}$ & $3.2^{+0.1}_{-0.1} \times 10^{14}$ & 41.9 \\
BCG-S & Hernquist &$4.36^{+0.09}_{-0.09} \times 10^{6}$ & $1.18^{+0.02}_{-0.02} \times 10^{3}$ & $3.2^{+0.1}_{-0.1} \times 10^{14}$ & 57.3 \\
Midpoint & NIS & $5.71^{+0.06}_{-0.05} \times 10^{7}$ & $1.32^{+0.01}_{-0.01} \times 10^{2}$ & $3.3^{+0.7}_{-0.7} \times 10^{14}$ & 204.5 \\
\midrule
\midrule
ACT0102 & Mass model & $\rho_0$ & $r_\mathrm{S}$ & $M(<500 \, \mathrm{kpc})$ & $\chi^2$ \\
\midrule
North clump & NIS & $4.42^{+0.12}_{-0.12} \times 10^{7}$ & $1.70^{+0.04}_{-0.03} \times 10^{2}$ & $4^{+1}_{-1} \times 10^{14}$ & 19.6\\
BCG-S & Hernquist & $1.74^{+0.14}_{-0.12} \times 10^{6}$ & $2.42^{+0.11}_{-0.12} \times 10^{3}$ & $3.6^{+0.4}_{-0.4} \times 10^{14}$ & 6.1\\
Midpoint & NIS & $1.19^{+0.05}_{-0.04} \times 10^{7}$ & $4.15^{+0.12}_{-0.17} \times 10^{2}$ & $2.8^{+0.2}_{-0.2} \times 10^{14}$ & 292.6\\
\midrule
\bottomrule 
\end{tabular}
\tablefoot{We tested the alternative positions of the projected total mass profile centers, as listed in Section \ref{sec.robtests}, for the Hernquist model. Central mass densities are expressed in $\mathrm{M_\odot /kpc^3}$ and length scales in kpc, while the total mass values are expressed in  $\mathrm{M_\odot}$.}
\end{table*}

\begin{table*}
\centering
\caption{Results of the increasing radius tests.}
\label{tab.radincrtest}
\begin{tabular}{cccccc} 
\toprule
\midrule
M0329 & $\rho_0$ & $r_\mathrm{S}$ &  & $M(<500 \, \mathrm{kpc})$  & $\chi^2$\\
\midrule
R=$50\arcsec$  & $9.68^{+0.39}_{-0.58} \times 10^{6}$ & $8.00^{+0.33}_{-0.20} \times 10^{2}$ & & $5.42^{+0.47}_{-0.07} \times 10^{14}$ & 48.5\\
R=$150\arcsec$ & $9.82^{+0.37}_{-0.35} \times 10^{6}$ & $7.95^{+0.19}_{-0.19} \times 10^{2}$ & & $5.37^{+0.06}_{-0.06} \times 10^{14}$ & 38.2\\
R=$200\arcsec$ & $9.73^{+0.40}_{-0.35} \times 10^{6}$ & $8.00^{+0.19}_{-0.21} \times 10^{2}$ & & $5.28^{+0.05}_{-0.05} \times 10^{14}$ & 36.7\\
R=$250\arcsec$ & $9.60^{+0.39}_{-0.34} \times 10^{6}$ & $8.08^{+0.19}_{-0.21} \times 10^{2}$ & & $5.51^{+0.05}_{-0.07} \times 10^{14}$ & 37.2\\
R=$300\arcsec$ & $9.48^{+0.39}_{-0.36} \times 10^{6}$ & $8.14^{+0.21}_{-0.21} \times 10^{2}$ & & $5.52^{+0.08}_{-0.07} \times 10^{14}$ & 37.7\\
\midrule
\midrule
A2744 & $\rho_0$ & $r_\mathrm{S}$ & $\beta$ & $M(<500 \, \mathrm{kpc})$ & $\chi^2$\\
\midrule
NFW & $5.57^{+0.13}_{-0.14} \times 10^{5}$ & $7.83^{+0.14}_{-0.13} \times 10^{2}$ & - & $6.04^{+0.03}_{-0.04} \times 10^{14}$ & 51.8 \\
NFW-T3 & $6.35^{+0.24}_{-0.24} \times 10^{5}$ & $7.13^{+0.20}_{-0.19} \times 10^{2}$ & - & $5.91^{+0.08}_{-0.07} \times 10^{14}$ & 77.7 \\
NIS & $1.92^{+0.03}_{-0.03} \times 10^{8}$ & $6.29^{+0.08}_{-0.07} \times 10^{1}$ & - & $5.18^{+0.02}_{-0.03} \times 10^{14}$ & 183.36\\
NIS-T3 & $2.29^{+0.04}_{-0.04} \times 10^{8}$ & $5.51^{+0.07}_{-0.07} \times 10^{1}$ & - & $4.88^{+0.03}_{-0.03} \times 10^{14}$ & 179.2 \\
Hernquist & $3.95^{+0.11}_{-0.10} \times 10^{6}$ & $1.36^{+0.03}_{-0.03} \times 10^{3}$ & - & $6.00^{+0.04}_{-0.04} \times 10^{14}$ & 54.1 \\
Hernquist-T3 & $4.63^{+0.17}_{-0.15} \times 10^{6}$ & $1.21^{+0.03}_{-0.03} \times 10^{3}$ & - & $5.80^{+0.07}_{-0.08} \times 10^{14}$ & 78.2 \\ 
Beta & $1.19^{+0.39}_{-0.41} \times 10^{9}$ & $9.22^{+3.89}_{-1.93} \times 10^{0}$ & $0.495^{+0.009}_{-0.004}$ & $6.85^{+0.07}_{-0.15} \times 10^{14}$ & 76.3\\
Beta-T3 & $9.34^{+14.26}_{-5.57} \times 10^{9}$ & $1.66^{+1.61}_{-0.81} \times 10^{0}$ & $0.466^{+0.002}_{-0.002}$ & $7.89^{+0.15}_{-0.02} \times 10^{14}$ & 67.7 \\
\midrule
\bottomrule 
\end{tabular}
\tablefoot{For M0329, we test if the parameter values of the Hernquist model change significantly when we increase the maximum radius mentioned above. For A2744, we check if all the proposed models give results similar to those in Table \ref{tab.paramtot} (indicated with T3), when the maximum radius for the projected mass profiles is set to $190\arcsec$. Central mass densities are expressed in $\mathrm{M_\odot /kpc^3}$ and length scales in kpc, while the total mass values are expressed in  $\mathrm{M_\odot}$.}
\end{table*}

Although we do not repeat these tests for every cluster in our sample, we can bring further arguments in favor of the robustness of our measurements. First, among the galaxy clusters of the regular class, it would not be very useful to measure the projected mass profile with a center shifted from the original one, since for this kind of galaxy clusters there is a single, well-determined peak of the total mass density (e.g. Figure 2 in \citealt{Bonamigo_2018_RXCJ2248_M0416_M1206} for AS1063 and M1206). For all the non-regular clusters, instead, we tested the effects of the center shift, since the presence of multiple peaks of total mass density does not allow us to determine unequivocably the position for the cluster center. Second, the radii of the set of circles, in which the total mass is measured, are spaced logarithmically, in order to sample better the innermost regions of the clusters, where the strong lensing measurements are more robust. Hence, even if we do not perform the increasing radius test on every galaxy cluster, we expect that the model parameter values, measured by fitting on smaller regions, do not change significantly if the size of the region is increased. This conclusion holds true especially for the most regular galaxy clusters in our sample. The previous argument is supported by an additional third test. Specifically, we verify that when the models (computed with the parameters fitted in the regions that are showed in Figure \ref{fig.clusters}) are extended to larger radii, the reduced $\chi^2$ values do not vary significantly.

\section{Conclusions} \label{sec.conclusion}

In this paper we studied the galaxy clusters RXJ2129, A2744, AS1063, M1931, M0416, M1206, M0329, M2129, and ACT0102 (cfr. Table \ref{tab.clusters} and Figure \ref{fig.clusters}), which are nine well-studied massive galaxy clusters at intermediate redshifts ($0.2 < z < 0.9$). Their rich photometric and spectroscopic datasets, which mainly come from CLASH, HFF, and CLASH-VLT programs combined with MUSE measurements, allowed for a detailed study of their mass composition (\citealt{Caminha_2016_AS1063}; \citealt{Caminha_2017_M0416}; \citealt{Caminha_2017_M1206}; \citealt{Bonamigo_2017_M0416}; \citealt{Bonamigo_2018_RXCJ2248_M0416_M1206}; \citealt{Bergamini_2019_RXCJ2248_M0416_M1206}; \citealt{Caminha_2019_CLASHSLmass}; \citealt{Caminha_2023_ElGordoSL}; \citealt{Bergamini_2023_A2744SL}), and the measurements of their cumulative radial total mass profiles. We performed the following analysis:

\begin{itemize}
\item We selected simple one-component spherically symmetric models (see Section \ref{sec.massmodels}), with 2 or 3 free parameters that can be simply related to the characteristic quantities of galaxy clusters, in order to fit their projected total mass profiles. The numeric fits are performed with a Monte Carlo Bayesian analysis of the posterior probability distribution of the free parameters of the models (Section \ref{sec.projMCMC}).

\item The results of the MCMC fits (Figure \ref{fig.totprof}) show that the NFW, the Hernquist, and the beta model profiles are generally good to reproduce the projected mass profiles of the selected galaxy clusters. Although the best mass model varies from cluster to cluster, the NIS does not figure in any case among them.

\item We employ the results on the projected total mass profiles to derive new possible scaling relations. In particular, we compare for each galaxy cluster the total mass values of the Hernquist model ($M_\mathrm{H}^\mathrm{tot} = M_\mathrm{H} (r\rightarrow + \infty)$) with the $M_{200\mathrm{c}}$ values measured from WL analyses. We also compare the scale lengths $r_\mathrm{S}$ of the Hernquist model and the $R_{200\mathrm{c}}$ value for each galaxy cluster, looking for a possible relation.

\item The comparison between SL-based ($M_\mathrm{H}^\mathrm{tot}$, $r_\mathrm{S}$) and WL-based measurements ($M_{200\mathrm{c}}$, $R_{200\mathrm{c}}$) leads to different scaling relations. We find a linear scaling law both for $M_\mathrm{H}^\mathrm{tot}$-$M_{200\mathrm{c}}$ and $r_\mathrm{S}$-$R_{200\mathrm{c}}$, while we test a power law dependence for the $M_\mathrm{H}^\mathrm{tot}$-$M_{200\mathrm{c}}$ relation (Figures \ref{fig.m200rel} and \ref{fig.R200rel}). We make this test on a reduced galaxy clusters sample, in which we discard the least regular clusters.

\item Although they are applied to very different regions of galaxy clusters, the correspondence between SL and WL measured $M_{200\mathrm{c}}$ (and $R_{200\mathrm{c}}$) values, in Figure \ref{fig.slwl200s}, shows that these two techniques can be linked under the proper conditions. This relation can be further investigated and exploited in future studies.  

\end{itemize}

We can draw three main conclusions about our modelization of galaxy clusters. First, there are different mass models which are able to reproduce equally well the projected, radial total mass profiles. In our galaxy cluster sample, we find that the best fitting models are the Navarro-Frenk-White (NFW, \citealt{NFW_1997_DMprofile}), the Hernquist \citep{Hernquist_1990_profile}, and the beta. Second, although galaxy clusters may present a complex structure, single-component spherical models are capable of reconstructing with very small relative errors ($\lesssim 5\%$) the inner mass distribution of the whole structure. Last, we find two scaling laws between the SL ($M_\mathrm{H}^\mathrm{tot}$, $r_\mathrm{S}$) and WL ($M_{200\mathrm{c}}$, $R_{200\mathrm{c}}$) measured quantities. The $M_\mathrm{H}^\mathrm{tot}$-$M_{200\mathrm{c}}$ and $r_\mathrm{S}$-$R_{200\mathrm{c}}$ relations can be used as powerful tools in future studies on the total mass distribution in galaxy clusters. This link is further enforced by the excellent matching between the SL and WL measured $M_{200\mathrm{c}}$ values, as well as between the $R_{200\mathrm{c}}$ values. 

The presented work provides new insights on the behavior of the total mass distribution in massive galaxy clusters. The study of the total mass profiles in galaxy clusters through well-known analytical models is surely to be enhanced, in order to have a fair balance between a simple description of such objects and an accurate tool for further measurements. As a first future development, we suggest to enlarge the sample of analyzed galaxy clusters: this would allow one to improve the scaling laws that we find in this work, in an extended effort to reduce the scatter of such relations. Through the modeling performed in this work, we set the basis for a sample of galaxy clusters with simple, but accurate, total mass models that can immediately render their most important properties. Moreover, a proper work on the quantities that are measurable through both SL and WL techniques could lead to surprising new methods for the study of galaxy clusters. 

\bibliographystyle{aa}
\bibliography{bibliografia}
\clearpage
\begin{appendix}

\onecolumn
\section{Parameter scaling with $M_{200\mathrm{c}}$} \label{sec.appendix}

\begin{figure*}
\includegraphics[width=\textwidth]{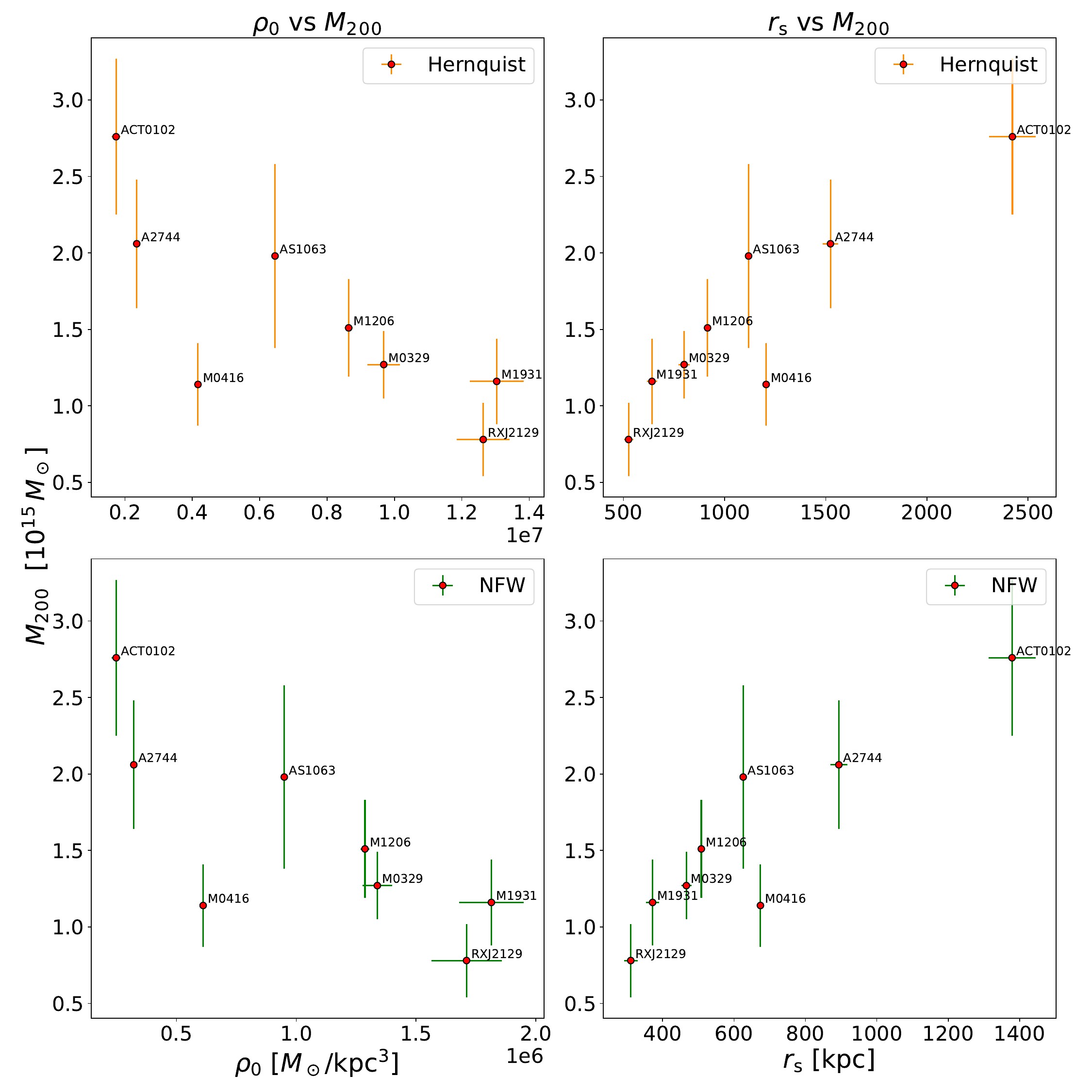}
\caption{Relation between the best-fitting parameters of the NFW and Hernquist profile for each galaxy cluster in the sample. On the left column we plot the $\rho_0$ vs $M_{200\mathrm{c}}$ relation, while on the right we plot $r_\mathrm{s}$ vs $M_{200\mathrm{c}}$.}
\label{fig.paramvsm200}
\end{figure*}

We plot in Figure \ref{fig.paramvsm200} the relation between the best-fitting parameters of the NFW and Hernquist profile for each galaxy cluster in the sample. We notice that the larger the scale radii $r_\mathrm{s}$ are, the higher are the corresponding $M_{200\mathrm{c}}$ values. This fact can be interpreted as a consequence of the concentration-mass relation (e.g. \citealt{Biviano_2017_concentrationrelation}). Regarding the $\rho_0$ values, we can justify their behavior by considering the anticorrelation between parameters that we discuss in Section \ref{sec.projeffects}. It is then straightforward to see that the most massive galaxy clusters are also those with the lowest scale density.

\end{appendix}

\end{document}